\documentclass[a4paper,11pt]{article}
\pdfoutput=1
\usepackage{jheppub}\usepackage[T1]{fontenc}

\usepackage[T1]{fontenc}
\usepackage[utf8]{inputenc}
\usepackage{lmodern, amsmath}
\usepackage{mathtools} 
\usepackage{cancel} 
\usepackage{MnSymbol} 
\usepackage{graphicx,adjustbox} 
\usepackage{mathrsfs} 
\usepackage{simpler-wick} 
\usepackage{comment}
\usepackage{autobreak}
\usepackage[noend]{algpseudocode}
\allowdisplaybreaks

\usepackage{hyperref}
\hypersetup{
	colorlinks=true,
	linktoc=page,
	citecolor=americanrose,
	linkcolor=cadmiumgreen,
	urlcolor=blue} 
\usepackage{xcolor}

\definecolor{americanrose}{rgb}{1.0, 0.01, 0.24}
\definecolor{cadmiumgreen}{rgb}{0.0, 0.42, 0.24}

\makeatletter
\newlength{\apb@width}
\newcommand{\autoparbox}[2][c]{\settowidth{\apb@width}{#2}\parbox[#1]{\apb@width}{#2}}
\newcommand{\includegraphicsbox}[2][]{\autoparbox{\includegraphics[#1]{#2}}}

\makeatletter\g@addto@macro\bfseries{\boldmath}\makeatother

\newcommand{\dd}{\mathrm{d}}
\newcommand{\hd}{\hat{\mathrm{d}}}

\newcommand{\hdelta}{\hat{\delta}}

\newcommand{\bu}{\bar{u}}
\newcommand{\bp}{\bar{p}}
\newcommand{\bM}{\bar{m}}

\newcommand{\cO}{\mathcal{O}}
\newcommand{\cW}{\mathcal{W}}
\newcommand{\cA}{\mathcal{A}}
\newcommand{\cI}{\mathcal{I}}

\newcommand{\dom}{w}
\newcommand{\gam}{\gamma}

\newcommand{\expval}[1]{\langle #1 \rangle}

\newcommand\scalemath[2]{\scalebox{#1}{\mbox{\ensuremath{\displaystyle #2}}}}

\newcommand{\oldornew}[1]{}

\title{\boldmath An Improved Framework for Computing Waveforms}

\author[a,b,c]{Giacomo Brunello}
\author[a]{Stefano De Angelis}

\affiliation[a]{Institut de Physique Théorique, CEA, CNRS, Université Paris-Saclay, F–91191 Gif-sur-Yvette cedex, France}
\affiliation[b]{Dipartimento di Fisica e Astronomia, Università degli Studi di Padova, Via Marzolo 8, I-35131 Padova, Italy.}
\affiliation[c]{INFN, Sezione di Padova,
Via Marzolo 8, I-35131 Padova, Italy.}

\emailAdd{giacomo.brunello@phd.unipd.it}
\emailAdd{stefano.de-angelis@ipht.fr}

\abstract{We combine the observable-based formalism (KMOC), the analytic properties of the scattering amplitude, generalised unitarity and the heavy-mass expansion with a newly introduced IBP reduction for Fourier integrals, to provide an efficient framework for computing scattering waveforms. We apply this framework to the scattering of two charged massive bodies in classical electrodynamics. Our work paves the way for the computation of the analytic one-loop waveform in General Relativity.}

\begin{document} 
\maketitle
\flushbottom


\section{Introduction}
\label{sec:introduction}

The emission of GWs emitted by a gravitating binary system has been studied for a long time in General Relativity, starting from the original leading-order quadrupole formula provided by Einstein \cite{Einstein:1918btx}, and from the pioneering computations of Kovacs and Thorne, and of Peters \cite{Peters:1970mx,Kovacs:1977uw,Kovacs:1978eu}. The Multipolar-Post-Minskowskian formalism (MPM) \cite{Blanchet:1985sp,Blanchet:1989ki,Blanchet:2004ek} has been one of the most effective methods for computing analytically the GWs emission from generic sources at very high precision for bound and unbound systems~\cite{Blanchet:2023bwj,Blanchet:2023sbv,Bini:2023fiz,Bini:2024rsy}. A novel framework to study the relativistic two-body problem emerged from particle physics and is based on effective field theories (EFTs) reasoning \cite{Goldberger:2004jt}. Classical bodies are studied in an EFT approach as point particles interacting via the exchange of gravitons, and their internal structure, such as spin and tidal forces, is encoded in non-minimal coupling to the gravitational field.

Classical in-in observables are directly related to scattering amplitudes via an observa\-ble-based approach (KMOC formalism) \cite{Kosower:2018adc}, reviewed in section \ref{sec:analyticity}. Within this framework, the waveform can be connected to the Fourier transform (FT) of $2\to 3$ scattering amplitudes \cite{Cristofoli:2021vyo}, after suitably changing some $i \epsilon$ prescriptions \cite{Caron-Huot:2023vxl}. 

The computation of the amplitudes is simplified by taking the classical expansion at the integrand level, before integrating over the loop momenta and taking the FT. Exploiting the method of regions \cite{Beneke:1997zp}, the classical regime can be identified with the so-called soft region \cite{Cheung:2018wkq,Bern:2019nnu,Bern:2019crd}, which corresponds to long-range interactions mediated by internal gravitons, where the loop momenta scales as the transferred and radiated momenta ($q$ and $k$, respectively): $\ell \sim q \sim k \ll m$. Alternatively, this classical limit can be performed as a heavy-mass expansion \cite{Brandhuber:2021kpo,Brandhuber:2021eyq}. The construction of the integrand can be further simplified by borrowing tools from generalised unitarity \cite{Bern:1994zx,Bern:1994cg,Bern:1995db,Britto:2004nc,Britto:2005ha,Britto:2006sj,Anastasiou:2006jv,Ossola:2006us,Anastasiou:2006gt}.
The integration instead is made possible by powerful methods developed in precision physics for multi-loop computations, such as integration-by-parts identities  (IBPs) \cite{Tkachov:1981wb,Chetyrkin:1981qh,Laporta:2000dsw,Smirnov:2008iw,Maierhofer:2017gsa,Lee:2012cn} -- which allow reducing all Feynman integrals (FIs) for a process in terms of a minimal set of linearly-independent integrals, known as Master Integrals (MIs) --,\footnote{Many computer programs have been developed to efficiently generate and solve IBPs \cite{Anastasiou:2004vj,Smirnov:2008iw,vonManteuffel:2012np,Lee:2012cn,Maierhofer:2017gsa,Wu:2023upw}. Throughout this project, we have been working mainly with \texttt{LiteRed}~\cite{Lee:2012cn}.} and differential equations (DEs) 
\cite{Kotikov:1990kg,Remiddi:1997ny,Gehrmann:1999as,Argeri:2007up,Henn:2013pwa,Henn:2014qga,Argeri:2014qva,Lee:2014ioa}.
Using these tools, the waveform has been computed at tree level for spinless particles using worldline formalism \cite{Jakobsen:2021smu,Mougiakakos:2021ckm} and including spin corrections \cite{Jakobsen:2021lvp,DeAngelis:2023lvf,Brandhuber:2023hhl,Aoude:2023dui}. The computation has been extended to the one-loop order and spinless bodies \cite{Brandhuber:2023hhy,Herderschee:2023fxh,Georgoudis:2023lgf,Elkhidir:2023dco,Caron-Huot:2023vxl}, and including linear-in-spin corrections \cite{Bohnenblust:2023qmy}. 
It has recently found agreement with Multipolar Post-Minkowskian (MPM) formalism in the small velocity expansion \cite{Bini:2023fiz,Georgoudis:2023eke,Bini:2024rsy,Georgoudis:2024pdz}. 
The one-loop amplitude contains many terms and spurious poles \cite{Brandhuber:2023hhy,Herderschee:2023fxh,Bohnenblust:2023qmy} (Gram determinant poles appearing from the reduction of tensor integrals), making it hard to think of an analytic evaluation of the FT -- for example, the small-velocity expansion was only possible via a cumbersome reorganisation of the different terms appearing \cite{Bini:2024rsy}.

The main goal of this work is to provide an efficient framework to compute the analytic waveform, treating the FT and the loop integrals as a genuine two-loop integration. The study of the analytic and algebraic properties of these Fourier-loop (FL) integrals will allow us to avoid the appearance of these poles in intermediate steps. Improvements can be made on several levels.
\begin{enumerate}
    \item \textit{Integrand.} From the analytic properties of the FL integrals, it is possible to select only the contributions to the amplitude giving rise to long-range interactions in the classical limit, discarding all contact terms. In the classical regime, only singularities related to internal on-shell gravitons survive. When we consider the Fourier integral of a function, we can exploit Cauchy's theorem to select these contributions, which can be treated as generalised unitarity cuts \cite{Kosower:2011ty,Badger:2012dv,Mastrolia:2012an,Ita:2015tya,Bourjaily:2015jna}. The building blocks fed into the cuts are computed in the heavy-mass expansions. 
    \item \textit{Spurious poles.} We perform the tensor decomposition at the level of the full FL integral, thus bypassing the problem of spurious poles.
    \item \textit{Integral reduction.} FIs of a given family live in a finite-dimensional vector space and the coefficients of the IBP decomposition of a given FI into a basis can be obtained from their intersection numbers \cite{Mastrolia:2018uzb,Frellesvig:2019uqt}. Integrals containing an exponential factor such as FTs can be analysed in the same spirit: this has been first seen in the context of confluent hypergeometric functions \cite{Matsumoto1998-2,majima2000}, and recently in relevant physical applications \cite{Luo:2019szz,Cacciatori:2022mbi,Brunello:2023fef}. Hence, Fourier integrals can be decomposed into a basis of MIs. This will leave us with the evaluation of a few integrals, which are the FTs of the scalar one-loop MIs. 
\end{enumerate}
We evaluate for the first time the analytic one-loop waveform in classical electrodynamics for arbitrary velocities, paving the way for the analogous calculation in General Relativity, that will be presented elsewhere \cite{Brunello:toappear}.

\section{Classical observables and Analyticity beyond the Physical Region}
\label{sec:analyticity}

\subsection{KMOC and the Waveform}

We start by reviewing the \textit{observable-based formalism} (KMOC) \cite{Kosower:2018adc,Cristofoli:2021vyo,Caron-Huot:2023vxl} for computing (classical) in-in observables from QFT scattering amplitudes. In general, we are interested in computing asymptotic observables, \textit{i.e.} the expectation value of some observable in the far future, after preparing the initial state of the system $| \psi \rangle_{\rm in}$ in the far past. We let the state transform with time and the evolution from the far past to the far future is given by the unitarity $S$ operator. We are generally interested in measuring the variation of the system: 
\begin{equation}
	\Delta\expval{\cO} = \expval{\cO}_{\rm out} - \expval{\cO}_{\rm in} =\, _{\rm out}\langle \psi | \cO | \psi\rangle_{\rm out} - \, _{\rm in}\langle \psi | \cO | \psi\rangle_{\rm in} =\, _{\rm in}\langle \psi | S^\dagger [\cO, S] | \psi\rangle_{\rm in}\ ,
\end{equation}
with
\begin{equation}
	S^\dagger S = \mathbf{1}\ .
\end{equation}

The KMOC formalism has been developed in the context of the (classical) two-body problem in gravity and electrodynamics. In this case, we consider $| \psi \rangle_{\rm in}$ to be a two-particle state, integrated against an on-shell wavefunction. Then, the initial state is\footnote{Most of the time, we will work in natural units $\hbar = c = 1$ and we will use the mostly negative metric signature.}
\begin{equation}
	| \psi \rangle_{\rm in} = \int\!\! \dd\Phi(p_1) \dd\Phi(p_2) \, \phi_1(p_1) \phi_2(p_2)\, e^{i (b_1 \cdot p_1 + b_2 \cdot p_2)}\, | p_1, p_2 \rangle_{\rm in}\ ,
\end{equation}
where $\dd \Phi (p_i)$ is the Lorentz-invariant on-shell phase-space (LIPS) measure\footnote{Here and in the rest of the paper, we follow the \textit{hat} notation introduced in the original KMOC paper~\cite{Kosower:2018adc}, \textit{i.e.} $\hd^D x = \frac{\dd^D x}{(2\pi)^D}$ and $\hdelta^D (x) = (2\pi)^D \delta^D (x)$.}
\begin{equation}
	\dd \Phi (p_i) = \hd^D p_i \, \Theta(p_i^0)\, \hdelta(p_i^2 - m_i^2)\ ,
\end{equation}
and $| p_1, p_2 \rangle_{\rm in}$ is a state constructed from two-particle momentum eigenstates, with wavepackets $\phi (p_i)$, which are well separated by an impact parameter $b^\mu = b_1^\mu - b_2^\mu$.\footnote{So far, the classical limit would enter only in the explicit form of the $\phi_i(p_i)$, as choosing
\begin{equation}
	\phi(p_1,p_2) = \phi_1(p_1) \phi_2(p_2)\, e^{i (b_1 \cdot p_1 + b_2 \cdot p_2)}\ ,
\end{equation}
is simply stating that we are not putting any effort into preparing the system in an entangled state.}\footnote{From now on, we will suppress the subscript ``in''.}

Then, we find
\begin{equation}
	\label{eq:quantum_KMOC_exp_value}
	\Delta \expval{\cO} = \!\int\! \prod_{i=1}^{4}\dd \Phi(p_i) \ \phi_1(p_1) \phi_2(p_2) \phi_2(p_1^\prime)^* \phi_1(p_2^\prime)^*\
	e^{i b_1 \cdot\left(p_1-p_1^\prime\right) + i b_2 \cdot (p_2 - p_2^\prime)} \
	\langle p_1^\prime, p_2^\prime | S^\dagger [\mathcal{O}, S] | p_1, p_2 \rangle \ .
\end{equation}
At this point, we can rewrite $p_i^\prime = p_i - q_i$.
The first instance of the \textit{classical limit} enters in the assumption that the Compton wavelength of the external particles is the smallest length scale in the problem. In particular, it must be much smaller than the characteristic spread of the wavepacket (for a detailed discussion on this point, we refer to Section~4 and Appendix~B of the original KMOC paper~\cite{Kosower:2018adc}). In other words, the wavepackets $\phi_i (p_i)$ are sharply peaked around the classical value of the two incoming momenta and any (quantum) deviation is \textit{exponentially suppressed}:\footnote{The wavepackets should be more properly thought as functions of the quantum momenta $p_i^\mu$, some classical momentum $\hat{p}_i^\mu$ and a bunch of characteristic scales describing the internal structure of classical body we are considering (spin, tidal deformability, etc.) which we are ignoring in this work.}
\begin{equation}
	\phi_i (p_i^\prime) \approx \phi_i (p_i)\cdot (1 + \text{quantum corrections})\ .
\end{equation}
Then, assuming that the wavefunctions are properly normalised ($\int \!\dd \Phi(p_i) |\phi_i(p_i)|^2 = 1$), the classical expectation value becomes to first approximation insensitive to the details of the wavepackets:
\begin{equation}
	\label{eq:classical_KMOC_exp_value}
	\Delta \expval{\cO}= \!\int\! \dd \mu \ \langle \bp_1-   \frac{q_1}{2}, \bp_2 - \frac{q_2}{2} | S^\dagger [\mathcal{O}, S] | \bp_1 + \frac{q_1}{2}, \bp_2 + \frac{q_2}{2} \rangle\ ,
\end{equation}
where
\begin{equation}
\label{eq:on_shell_measure}
	\dd \mu = \prod_{i=1}^{2}\hd^D q_i \, \hdelta(-2 \bp_i \cdot q_i)\, e^{i b_i \cdot q_i}
\end{equation} 
and $\bp_i = p_i - \frac{q_i}{2}$ are the well-known \textit{barred} variables, and we have not made any expansion in the argument of the delta functions. Thus, the observables can be generically expressed as a $(D-2)$-dimensional  -- as $q_1^\mu$ and $q_2^\mu$ are related by momentum conservation -- FT of an in-in correlator. The latter can be further expressed in terms of the scattering amplitudes.\footnote{Reference~\cite{Caron-Huot:2023ikn} argues that it can even obtained from the scattering amplitudes after analytic continuation, before taking the classical limit.} In particular, we can always write 
\begin{equation}
	S = \mathbf{1} + i T\ ,
\end{equation}
and the scattering amplitudes are usually defined as the transition element of $T$, modulo the momentum-conserving delta function. Following \cite{Herrmann:2021tct}, we can write 
\begin{equation}
	\Delta \expval{\cO}= \!\int\! \dd \mu \left(\mathcal{I}_{O, \mathrm{v}}+\mathcal{I}_{O, \mathrm{r}}\right)\ ,
\end{equation}
where
\begin{equation}
	\begin{split}
		\label{eq:HPRZ_split}
		\mathcal{I}_{O, \mathrm{v}}+\mathcal{I}_{O, \mathrm{r}}
		=&\ i\, \langle p_1^\prime, p_2^\prime |[\mathcal{O}, T]| p_1, p_2\rangle+\langle p_1^\prime, p_2^\prime |T^{\dagger}[\mathcal{O}, T]| p_1, p_2\rangle \ ,
	\end{split}
\end{equation}
\textit{i.e.} the integrand of \eqref{eq:quantum_KMOC_exp_value} has been split into \textit{virtual} and \textit{real} contributions, respectively. The cluster decomposition principle suggests that any observable $\cO$ can be written as a sum of products of annihilation and creation operators (for detailed discussion on this point, see the original paper~\cite{Wichmann:1963aba} or Chapter~4 of Weinberg's book~\cite{Weinberg:1995mt}), which implies that the former can be written in terms of scattering amplitudes and the latter as integrated products of them (inserting a complete set of states between the operator products).

In this paper we will focus on a specific class of observables, the \textit{waveforms} in classical electrodynamics, which correspond to the value of the electric field in the far future as a function of the (retarded) time and the angles on the celestial sphere. The operator we will consider is
\begin{align}
	\cW_{ED} &= \varepsilon_{h}^\mu A_\mu\ ,
\end{align}
where $h$ stands for the helicity configuration of the waves which we are interested in ``measuring'' and $A_\mu$ is the electromagnetic field. Indeed, we know that the expectation value of the field itself $\expval{A^{\mu}(x)}$ is not observable -- it is not a gauge-invariant quantity: indeed, gauge transformations change its value (we have in mind transformations which vanishes at infinity, \textit{i.e.} we are not taking into account large gauge). On the other hand, when we consider its behaviour in the light-like future $\mathscr{I}^+$, the LSZ reduction sets to zero all the non-linearities. Then, the $\varepsilon_h^{\mu}$ plays the role of a projector onto the physical -- gauge-invariant -- states of the field, ensuring that $\varepsilon_h^{\mu}\expval{A_{\mu}(u)}|_{\mathscr{I}^+}$ satisfies Ward identities. Analogous statements can be made for the metric (and for radiated scalars, as well), considering diffeomorphisms (or field redefinitions, in general) \cite{Bini:2024rsy}. The computation of the analytic gravitational waveform at one-loop will be presented elsewhere \cite{Brunello:toappear}.

With a proper choice of the normalisation for the polarisations,\footnote{The proper normalisation being $\varepsilon_h(-k) = \varepsilon_{-h}(k) = \varepsilon^*_h(k)$ and $\varepsilon_h \cdot \varepsilon_{h^\prime} = - \delta_{h,-h^\prime}$.} the waveform operator can be written in terms of a single creation or annihilation operator
\begin{equation}
	\cW_h(x) = \!\int\! \hd \Phi(k) \left[ e^{- i k \cdot x} a_{-h} (k) + e^{+ i k \cdot x} a_{h}^\dagger (k)\right] \ .
\end{equation}
where $\dd \Phi (k)$ is the massless Lorentz-invariant on-shell phase-space (LIPS) measure. For $x^0 = t \to + \infty$ and $|\vec{x}| \coloneq r \to +\infty$, the integral can be evaluated using the saddle-point approximation:
\begin{equation}
\begin{split}
	\Delta \expval{\cW_h }(u,\vec{n}) = -\frac{i}{4\pi r} \!\int_0^\infty\! \hd \omega \, \!\int\! \dd \mu\ & e^{- i \omega u} \langle p_1^\prime, p_2^\prime |S^{\dagger}[a_{-h}(k), S]| p_1, p_2\rangle\\
	- & e^{+i \omega u} \langle p_1^\prime, p_2^\prime |S^{\dagger}[a_{+h}^\dagger(k), S]| p_1, p_2\rangle\ ,
\end{split}
\end{equation}
where $u=t-r$ is the \textit{retarded time} and $k^\mu = \omega n^{\mu} = \omega (1,\vec{n})$, with $\vec{n}$ a unit vector.
Finally, we can write the waveform in terms of the relevant scattering amplitudes:
\begin{equation}
\label{eq:waveform}
\begin{split}
	\Delta \expval{\cW_h }(u,\vec{n}) = & \frac{1}{4\pi r} \!\int_0^\infty\! \hd \omega \, \!\int\! \dd \mu\, \times\\
	\Big\{ & \hdelta^D (q_1+q_2 - k) e^{- i \omega u}\, \left[\cA(p_1 p_2\to p_1^\prime p_2^\prime k^{-h}) -i \cA(p_1 p_2 \to X k^{-h}) \otimes \cA^* (p_1^\prime p_2^\prime \to X)\right]\\
	+ & \hdelta^D (q_1+q_2 + k) e^{i \omega u}\, \left[\cA^* (p_1^\prime p_2^\prime \to p_1 p_2 k^{+h}) +i \cA(p_1 p_2 \to X) \otimes \cA^* (p_1^\prime p_2^\prime \to X k^{+h})\right] \Big\}\ ,
\end{split}
\end{equation}
where $\cA(p_1 p_2\to p_1^\prime p_2^\prime k^{-h}) = \langle p_1^\prime, p_2^\prime, k^{-h} |T| p_1, p_2\rangle$, $\otimes$ is understood as a completeness resolution of the identity build out the in states $|X\rangle$.\footnote{For example, in the classical limit we have
\begin{equation}
\begin{split}
	\cA^* (p_1^\prime p_2^\prime \to X)\otimes \cA(p_1 p_2 \to X k^{-h}) = \sum_{n} \!\int\! \dd \Phi(p_1^{\prime \prime}) \dd \Phi(p_2^{\prime \prime}) \prod_{i=1}^n \dd \Phi(k_i) \sum_{h_i} & \langle p_1^\prime p_2^\prime |T^{\dagger} | p_1^{\prime \prime} p_2^{\prime \prime} k_1^{h_1}\dots k_n^{h_n}\rangle \times \\
	\times & \langle p_1^{\prime \prime} p_2^{\prime \prime} k_1^{h_1}\dots k_n^{h_n} k^{-h} | T| p_1 p_2\rangle\ .
\end{split}
\end{equation}
} 
The second line ensures that the waveform is real, after stripping off the polarisations. Then, we are going to ignore this term for the moment. Reference~\cite{Caron-Huot:2023vxl} showed that the second terms in both lines are needed to restore the correct causality properties: the waveform is an in-in observable, while amplitudes are in-out observables. Like in classical physics, we fix initial boundary conditions (represented here by $|\psi \rangle_{\rm in}$) and we let the system evolve. This suggests that a causal evolution of the system requires \textit{retarded} propagators, while in the amplitude we have Feynman-$i\epsilon$'s. The role of the terms which are quadratic in the scattering amplitude is changing the $i\epsilon$ prescription from Feynman to retarded. Moreover, this term is needed to subtract from the amplitude iterations of lower order in perturbation theory, which are usually referred to as \textit{super-classical} or \textit{classically singular} contributions. For this reason, we will refer to this term as KMOC \textit{subtraction}. On the other hand, since \textit{generalised unitarity} is insensitive to the prescriptions for the propagators, we are also going to ignore this subtraction until the very last step, \textit{i.e.} until the integral evaluation.

\subsection{Analytic properties of the five-point amplitude and the Fourier transform}

In the previous section, we showed that classical observables are related to amplitude by a FT, after a suitable change of $i \epsilon$ prescriptions (and taking the classical limit). In this section, combining the basic properties of the FTs and the properties of Feynman integrals, we are going to explain how to simplify the analytic computation of waveforms, which has been a challenging problem away from the non-relativistic limit. Our method is the generalisation of the strategy presented in \cite{DeAngelis:2023lvf}, beyond the leading-order approximation.

Cauchy's theorem tells us that the (inverse) FT (FT) of a function is fully determined by the analytic structure of its analytic continuation on the (upper) lower-half complex plane. While the FT for the fully quantum KMOC expectation value~\eqref{eq:quantum_KMOC_exp_value} is expected to be absolutely convergent, the classical limit is only well-defined within the space of tempered distributions, with several contributions which are localised in impact parameter space (IPS) -- they are proportional to $\delta(\sqrt{-b^2})$ and its derivatives. On the other hand, the classical limit corresponds to the leading long-range term and contact interactions are not computed within this approximation. Then, it would be desirable to select only terms which contribute to long-range interactions and discard the rest. In this section, we will exploit the analytic structure of the scattering amplitudes on the physical sheet (\textit{i.e.} in complex kinematics which are continuously connected to the real kinematics -- \textit{physical region} -- without crossing any branch cut) to isolate such terms. In particular, we will show how distributional contributions appear from the classical limit, which pushes some of the singularities of the quantum amplitude to infinity in the complex $q_{1,2}^2$ planes.

A second unpleasant feature of generic higher-point scattering amplitudes and, in particular, those in the integrand of equation~\eqref{eq:waveform} is the appearance of spurious poles, which are singularities of the amplitude beyond the physical sheet, but are completely smooth in the principal sheet. These appear as higher-order poles in the rational coefficient of the transcendental functions and the cancellations in the physical region are highly non-trivial. The origin of these singularities stems from the tensor Feynman integrals and, therefore, they correspond to Gram determinants in various dimensions. Since on the physical sheet there are no unphysical singularities, complex contour deformation can be performed smoothly through them.\footnote{This fact has been used in reference~\cite{Herderschee:2023fxh} to perform the FT numerically, as the presence of these poles in the physical region makes the numerical stability very hard to handle. A  stable numerical evaluation of the full amplitude requires working with high-precision numerics.} This information, combined with a Passarino-Veltman~\cite{Passarino:1978jh} reduction at the level of the combined FT and loop integration, allowed us to bypass this problem.

\begin{figure}[!t]
    \centering
    \includegraphicsbox[scale=.9]{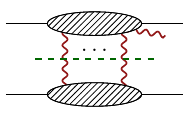} \>
    \includegraphicsbox[scale=.9]{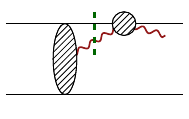} \>
    \caption{The unitarity cuts associated to the singularities on the physical sheet of the (full-quantum) five-point amplitude, up to permutations. Indeed, we have two unitarity cuts of the first type and four of the second type (which are pushed to infinity in the $q_i^2$-complex plane in the classical limit). We should emphasise that the first diagram corresponds to singularities which are not in the physical region but can only be probed within complex kinematics: \textit{i.e.} the graviton propagators should not be thought as delta function, but as residues on their poles (they are commonly referred to as \textit{generalised unitarity cuts}).}
	\label{fig:cuts}
\end{figure}

\begin{figure}[!th]
    \centering
    \includegraphics{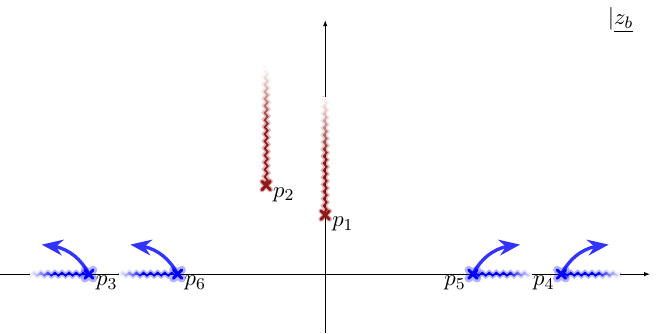}
    \caption{The singularities (poles at tree-level and branch points at loop) of the quantum five-point amplitude. As we take the classical limit (formally, $m_i\to +\infty$), the blue singularities -- right diagram in Figure~\ref{fig:cuts} -- are pushed to infinity and we are left only with the red singularities -- left diagram in Figure~\ref{fig:cuts} -- in complex kinematics corresponding to $q_1^2=0$ or $q_2^2=0$ (even though we are not displaying them, in the $z_{v,q}$ complex plane there are two symmetrical singularities in the lower-half plane). In terms of generalised unitarity cuts, terms which are not probed by the cuts in the $q_i^2$ channels (but only by Compton-like cuts, on the RHS of Figure~\ref{fig:cuts}) do not contribute to the classical waveforms.}
    \label{fig:poles}
\end{figure}

We can start from the waveform in the frequency domain in terms of the scattering amplitudes
\begin{equation}
\label{eq:waveform_frequency_space}
\begin{split}
	\Delta \expval{\cW_h }(\omega,\vec{n}) & = \frac{1}{4\pi r} \!\int\! \prod_{i=1}^{2}\hd^D q_i \, \hdelta(-2 \bp_i \cdot q_i)\, e^{i b_i \cdot q_i} \, \hdelta^D (q_1+q_2 - k) \cA_5(p_1 p_2\to p_1^\prime p_2^\prime k^{-h}) + \dots\ ,\\
	& = \frac{e^{i \omega\, n \cdot b_2}}{4\pi r} \!\int\! \hd^D q \, \hdelta(2 \bp_1 \cdot q) \hdelta(2 \bp_2 \cdot (k-q))\, e^{i b \cdot q} \, \cA_5(p_1 p_2\to p_1^\prime p_2^\prime k^{-h}) + \dots\ ,
\end{split}
\end{equation}
where the dots stand for the KMOC subtraction and $b^\mu = b_1^\mu - b_2^\mu$. It is convenient at this point to use a $D$-dimensional generalisation of the $q$-integral parametrisation introduced in reference~\cite{Cristofoli:2021vyo}:
\begin{equation}
	\label{eq:qparam}
	q^\mu= z_1 \bu_1^\mu + z_2 \bu_2^\mu + z_b \tilde{b}^\mu + z_v v^\mu\, ,
\end{equation}
where
\begin{align}
	\bu_1^\mu= \frac{\bp_1^\mu}{\bM_1}\, ,\quad  \bu_2^\mu= \frac{\bp_2^\mu}{\bM_2}\, , \quad\tilde{b}^\mu = \frac{b^\mu}{\sqrt{-b^2}} \, , \quad v^2 = - 1.
\end{align}
$v^\mu$ is a $D-3$ dimensional unit vector, which is orthogonal to $\bu_1$, $\bu_2$ and $\tilde{b}$. $b$ is the asymptotic impact parameter, then we also have $b\cdot \bu_1 = b\cdot \bu_2 =0$ ($\bu_i^2 = 1$ and $\bp_i^2 = \bM_i^2$).
The resulting Jacobian is
\begin{equation}
	d^D q = \sqrt{\gamma^2-1}\, z_v^{D-4} \, d^2 z\, d^{D-4}v\ ,
\end{equation}
with $\gamma=\bu_1\cdot \bu_2$, such that \eqref{eq:waveform_frequency_space} becomes
\begin{equation}
\label{eq:Davids_FT_parametrisation}
	\Delta \expval{\cW_h }(\omega,\vec{n})= \frac{e^{i \omega n \cdot b_2}}{(4 \bM_1 \bM_2) \sqrt{\gamma^2-1}} \int\!\! \hd^{D-4}v\, \hd^2 z\, z_v^{D-4} e^{ - i z_b \sqrt{-b^2}}\, \left.\left(\cA_5 + \dots \right)\right|_{ z_1 =  \frac{\gamma\, w_2}{\gamma^2-1} , \, z_2 = -  \frac{w_2}{\gamma^2-1} }\ ,
\end{equation}
where $\dd^{D-4}v$ is an angular integration over $(D-4)$-dimensional sphere, $z_b$ are integrated along the real axis and $z_v$ over the positive real axis. For $D=4$, this parametrisation recovers the one presented in Appendix~C~of~\cite{Cristofoli:2021vyo}, with the integration over $z_v$ to be performed along the full real axis (no angular integration has to be performed) and $v^\mu = \frac{\epsilon^{\mu \nu \rho \sigma} \bu_{1 \nu}  \bu_{2 \rho} b_{\sigma}}{\sqrt{\gamma^2-1}}$, fixed. Moreover, we have also defined the kinematic variable
\begin{equation}
	w_i = \bu_i \cdot k \ .
\end{equation}
The five-point scattering amplitude is a function of five Mandelstam invariants and the physical kinematics is defined by
\begin{equation}
		\gamma > 1\ ,\qquad w_i > 0\ ,\qquad -q_i^2 > 0 \ .
  \label{eq:kinematics}
\end{equation}
It is convenient to express $q_{1,2}^2$ in terms of $z_v$, $z_b$ and the other variables:
\begin{equation}
	\begin{split}
		- q_1^2 &= z_b^2+z_v^2+\frac{w_2^2}{\gamma ^2-1}\ ,\\
		- q_2^2 &= z_b^2+2 z_b \tilde{b}\cdot k+z_v^2+2 z_v k\cdot v-\frac{w_2 \left(w_2-2 \gamma  w_1\right)}{\gamma ^2-1}\ .
	\end{split}
\end{equation}
The classical five-point amplitude (and its quantum parent) has singularities at $q_i^2=0$ (poles at tree level and branch points at loop) corresponding to intermediate gravitons going on-shell. The quantum amplitude has additional singularities in the $z_{b,v}$ complex plane, corresponding to the massive (classical) particles going on-shell:
\begin{equation}
	\begin{split}
		(\bp_1 \pm \frac{q_1}{2} - k)^2 - m_1^2 &= - 2 \bM_1 w_1 \mp \frac{q_1^2 - q_2^2}{2}\ ,\\
		(\bp_2 \pm \frac{q_2}{2} - k)^2 - m_2^2 &= - 2 \bM_2 w_2 \pm \frac{q_1^2 - q_2^2}{2}\ .
	\end{split}
\end{equation}
The unitarity cuts associated with these singularities are shown in Figure~\ref{fig:cuts}. A striking fact about such points is that, taking the classical limit (for example, in the form of \textit{heavy-mass} limit \cite{Brandhuber:2021eyq}), the singularity is pushed to infinity, as shown in Figure~\ref{fig:poles}. We should emphasise here that the singularities of the five-point amplitude are the same as those of the KMOC integrand (which is an in-in correlator), as the $i\epsilon$ prescriptions do not modify the branch points -- they change how the branch cuts are approached. If we consider the FT after taking the classical limit, the point at infinity is not regular and the integral is not uniformly convergent (as we may have terms which are polynomial in $q_1^2 - q_2^2$). On the other hand, such contributions integrate to contact interactions $\delta (\sqrt{-b^2})$ (or its derivatives) in the IPS. Moreover, to keep the leading long-range -- classical -- contributions, we are going to consider only the leading order in the heavy-mass expansion in the integrand representation of the amplitude.\footnote{This corresponds to taking into account only the first term in the \textit{soft region} expansion, which is by definition polynomial in the masses. The \textit{hard region} will give terms which are analytic in $q_i^2$ and transcendental in the masses. This analysis has been carried out in detail in reference~\cite{Caron-Huot:2023vxl}.}
 
\section{The soft expansion and the integrand from generalised unitarity}
\label{sec:unitarity}

In this section, we are going to present a simplified strategy to construct the integrand for the waveform. It is important to stress that the integrand is in general a function of the loop momentum and the momentum mismatches $q_i^\mu$, and both such integrations will be discussed in the following section. The two main ingredients discussed in this section are
\begin{enumerate}
    \item \textit{generalised unitarity} at the level of the combined FT~\eqref{eq:waveform_frequency_space} and the loop integral,
    \item the asymptotic expansion in the \textit{soft regions} of the observable in the classical limit at the level of the generalised unitarity cuts, which was introduced in reference~\cite{Brandhuber:2021eyq} and it is referred to as the \textit{heavy-mass} expansion.
\end{enumerate}
In particular, generalised unitarity is important to isolate those terms which give non-trivial contributions to the FT. Indeed, we know that the FT is sensitive only to the singularities at $q_1^2=0$ and $q_2^2=0$ in the classical limit, while analytic terms would only give localised -- short-range -- contributions. As mentioned in the previous section, such singularities are probed by the unitarity cuts in Figure~\ref{fig:cuts} (left) and terms which vanish when probing any of these cuts can be discarded.

Then, in this section, we will start computing the Compton amplitude at tree level and one loop and the three-photon amplitude at tree level in the heavy-mass expansion, which appears on the two sides of the $q_i^2$-cuts. We are going to provide the result in a manifestly gauge-invariant form, similar to~\cite{Brandhuber:2021kpo}. Next, we are going to present an ansatz for the waveform integrand at one-loop and compute it in the case of electrodynamics.

\subsection{QED amplitudes in the heavy-mass expansion -- Tree-level}

We compute the four- and five-point tree-level amplitudes from Feynman diagrams (with $+i\epsilon$ prescription). Before glueing them in the generalised unitarity cuts, we can take the classical limit in the form of the heavy-mass expansion $\bM \gg k_i, q$ ($q=\sum_i^n k_i$, the incoming massive scalar has momentum $p$ and the outgoing has momentum $p-q$). Here, it is important to emphasise that the heavy-mass expansion has to be taken around the barred variables $\bp$, rather than $p$, as enforced by the on-shell measure \eqref{eq:on_shell_measure} (this is important to not miss classical terms from the expansion of the delta function, which would look like $\frac{\hbar}{\hbar}$). The matter-photon coupling is normalised as
\begin{equation}
    \cA_3 (p;k) = 2 e\, \varepsilon_k \cdot p\ .
\end{equation}
The four-point amplitude in the heavy-mass expansion takes the form
\begin{equation}
\label{eq:tree_Compton_classical}
    \begin{split}
        \tilde{\cA}_4^0(\bM \bu;k_1,k_2) & = 2                                                                    
        i e^2 \bM\, \hdelta(k_1\cdot \bu)\, \varepsilon_{1}\cdot \bu\, \varepsilon_{2}\cdot \bu + 2 e^2 \frac{\bu\cdot F_{1}\cdot F_{2}\cdot \bu}{k_1\cdot \bu^2} \\
        & + 
        \frac{i e^2}{4 \bM} \hdelta^{\prime \prime} (k_1\cdot \bu) \, k_1\cdot F_{2}\cdot \bu\, k_2\cdot F_{1}\cdot \bu + \cO (\bM^{-2})\ .
    \end{split}
\end{equation}
where the dot $\cdot$ stands for a contraction of Lorentz indices, $F_i^{\mu \nu} = k_i^\mu \varepsilon_i^\nu - k_i^\nu \varepsilon_i^\mu$ and we are using the notation
\begin{equation}
    \tilde{\cA}_{2+n}^0(\bp;k_1,\dots ,k_n) = \cA_{2+n}^0(\bp+\frac{q}{2}
    ;k_1,\dots ,k_n)\ .
\end{equation}
The second term in equation~\eqref{eq:tree_Compton_classical} is genuinely classical and the linear propagator has to be interpreted with a (symmetric-in-time) principal-value prescription:
\begin{equation}
    \frac{1}{x} = \frac{1}{2x + i \epsilon} + \frac{1}{2x - i \epsilon}\ .
\end{equation}
Here we have given one additional order to the needed classical expansion for reasons that will be clarified below.
The five-point amplitude is\footnote{The third term in the expansion involves a product of principal values which is a subtle distribution \cite{Davies:1996gee}. Such terms have to be interpreted as principal value prescriptions after disentangling linearly dependent denominators using partial fraction identities.}
\begin{equation}
\label{eq:five_point}
    \begin{split}
        \tilde{\cA}_5^0(\bp;k_1,k_2,k_3) = & - 
        e^3 \bM\, \hdelta(k_1\cdot \bu)\, \hdelta(k_2\cdot \bu)\, \varepsilon_{1}\cdot \bu\, \varepsilon_{2}\cdot \bu\, \varepsilon_{3}\cdot \bu + \text{perm.'s}\\
        & + 2 
        i  e^3 \hdelta(k_1\cdot \bu)\, \varepsilon_{1}\cdot \bu \frac{\bu\cdot F_{2}\cdot F_{3}\cdot \bu}{(k_2\cdot \bu)^2} + \text{perm.'s}\\
        & + 2 e^3 \frac{q\cdot F_{1}\cdot \bu\, \bu\cdot F_{2}\cdot F_{3}\cdot \bu}{\bM (k_1\cdot \bu)^2\, k_2\cdot \bu\, k_3\cdot \bu}  + \text{perm.'s} + \dots + \cO(\bM^{-2})\ ,
    \end{split}
\end{equation}
where the dots stand for additional terms proportional to $\delta (k_i\cdot \bu) \delta^{\prime \prime} (k_j\cdot \bu)$, which are irrelevant for the computation of the waveform -- to the order considered (for example, such terms are relevant for the computation of the momentum kick at two loops). These terms have not been considered in the original papers on the kinematic-algebra approach to the heavy-mass effective theory~\cite{Brandhuber:2021kpo,Brandhuber:2021bsf} and we leave a systematic understanding of these structures for future works, ignoring them in this work as we are genuinely interested in $\omega\neq 0$. For QCD and gravity amplitudes in the classical limit, this decomposition have been presented in reference~\cite{Bjerrum-Bohr:2021wwt}. 

At this point, it is important to understand the tree-level contributions appearing in the KMOC subtractions. We will have the contributions both from factorised four- and five-point amplitudes, which take the form
\begin{equation}
    \begin{split}
        \cA_{3}^0(p;k_1)\, \hdelta(2 p \cdot k_1)\, \cA_{3}^0(p-k_1;k_2) & = 2 e^2 \bM\, \hdelta(k_1\cdot \bu)\, \varepsilon_{1}\cdot \bu\, \varepsilon_{2}\cdot \bu     \\
        & - e^2 \, \hdelta^\prime(k_1\cdot \bu)\, \bu\cdot F_{1}\cdot F_{2}\cdot \bu + \cO(\bM^{-1})\ ,
    \end{split}
\end{equation}
and
\begin{equation}
\label{eq:cut_five_point}
    \begin{split}
        \cA_{4}^0(p;k_1,k_2)\, \hdelta((p-k_1-k_2)^2-m^2)\, & \cA_{3}^0(p-k_1-k_2;k_3) = \dots + 2 e^3 \hdelta(k_3\cdot \bu)\, \varepsilon_{3}\cdot \bu \frac{\bu\cdot F_{1}\cdot F_{2}\cdot \bu}{(k_1\cdot \bu)^2}   \\
        -& e^3 \hdelta^\prime (\bu \cdot k_3) \frac{q \cdot F_3 \cdot \bu \, \bu \cdot F_1 \cdot F_2 \cdot \bu}{\bM (k_1 \cdot \bu)^2}                             \\
        -& e^3 \hdelta (\bu \cdot k_3) \varepsilon_3 \cdot \bu \frac{ q \cdot k_3 \, \bu \cdot F_1 \cdot F_2 \cdot \bu}{\bM (k_1 \cdot \bu)^3} \\
        -& e^3 \hdelta (\bu \cdot k_3) \varepsilon_3 \cdot \bu \frac{k_2\cdot F_1 \cdot \bu \, k_3\cdot F_2 \cdot \bu }{\bM (k_1\cdot \bu)^3}                      \\
        +& e^3 \hdelta (\bu \cdot k_3) \varepsilon_3 \cdot \bu \frac{k_3\cdot F_1 \cdot \bu \, k_1\cdot F_2 \cdot \bu }{\bM (k_1\cdot \bu)^3} + \cO(\bM^{-2}) \ ,
    \end{split}
\end{equation}
where we are ignoring again all the terms which have two delta functions. According to equation~\eqref{eq:waveform}, on the right side we should take the complex conjugate amplitude but, at this order, we have only the three-point coupling and the complex conjugation does not play any role. The factorised five-point amplitude can be easily obtained from the expansion of four-point to sub-classical order~\eqref{eq:tree_Compton_classical} (it is manifest that the expansion of the delta mixes different orders in the heavy-mass). This form of the expansion makes it manifest that equation~\eqref{eq:cut_five_point} will combine with the five-point~\eqref{eq:five_point} to change the $i \epsilon$ prescription of the relevant massive propagator.

\subsection{QED amplitudes in the heavy-mass expansion -- One-loop}
\label{sec:QED_compton}

In this section, we compute the full quantum one-loop Compton amplitude using generalised unitarity. First, we probe the singularity in the $(P-k_1)^2$ channel and we get the full amplitude from symmetry. Indeed, in electrodynamics and, in general, in QED without additional matter (or simply in QED at one loop), the discontinuity across the threshold in the $(k_1+k_2)^2$ is zero. For example, this is not true in gravity and, when we apply these techniques to the general-relativity waveform, we need to treat such contributions carefully, as we will explain in the following section. 
For this computation, we need the tree-level four-point amplitude:
\begin{equation}
    \cA_4^0 (m\, u;k_1,k_2) = -8 e^2\frac{m^2\, u\cdot F_{1}\cdot F_{2}\cdot u}{\left(-2 m k_1\cdot u+i\epsilon\right) \left(2 m k_1\cdot u-q^2+i\epsilon\right)}\ .
\end{equation}
The ansatz for the Compton amplitude at one-loop can be written as follows:
\begin{equation}
    \label{eq:ansatz_Compton}
    \begin{split}
        \cA_4^1 (p;k_1,k_2) & = d_{\mu \nu}\, j_{1,1,1,1}^{\mu \nu} + d_{\mu}\, j_{1,1,1,1}^{\mu} + d\, j_{1,1,1,1} + c_{\mu}\, j_{1,0,1,1}^{\mu} + c\, j_{1,0,1,1} + b\, j_{1,0,1,0} + \dots \\
        & + \tilde{d}_{\mu \nu}\, \tilde{j}_{1,1,1,1}^{\mu \nu} + \tilde{d}_{\mu}\, \tilde{j}_{1,1,1,1}^{\mu} + \tilde{d}\, \tilde{j}_{1,1,1,1} + \tilde{c}_{\mu}\, \tilde{j}_{1,0,1,1}^{\mu} + \tilde{c}\, \tilde{j}_{1,0,1,1} + \tilde{b}\, \tilde{j}_{1,0,1,0} + \dots \ ,
    \end{split}
\end{equation}
where the $b$'s, $c$'s and $d$'s are rational coefficients of the kinematic variables and the (tensor) Feynman integrals are defined as
\begin{equation}
    j_{a_1,a_2,a_3,a_4}^{\mu_1 \dots \mu_n} = \int\! \hd^D\ell\, \frac{\ell^{\mu_1}\, \cdots\,\ell^{\mu_n}}{[\ell^2]_+^{a_1} [(p-\ell)^2 - m^2]_+^{a_2} [(p-k_1-\ell)^2 - m^2]_+^{a_3}[(p-q-\ell)^2 - m^2]_+^{a_4}}\ ,
\end{equation}
with $[p]_+ = p + i\epsilon$. $\tilde{j}^{\mu_1 \dots \mu_n}_{a_1,a_2,a_3,a_4}$ is obtained from the previous equation by symmetry $k_1 \leftrightarrow k_2$. The powers of loop momenta in the numerators come from naïve derivative counting in QED. The loop momenta appearing in the numerator are contracted either with the field strengths or in scalar products which are not in the denominator. The dots in the ansatz~\eqref{eq:ansatz_Compton} stand for integrals which have only transcendental weights carried by the masses, \textit{i.e.} they are proportional to $(m^2)^{-\epsilon}$. Such contributions are irrelevant for the classical limit \cite{Bini:2021gat} and we are going to ignore them.

Now, we can consider the integrand and take the residue in complex kinematics on the 2-torus encircling the poles $\ell^2$ and $(p-k_1-\ell)^2$. Then, we can fix the rational coefficients by matching our ansatz to the product of two tree-level Compton amplitudes, summing over the internal photon states. Since the polarisation vectors appear only inside linearised field strengths, we can use the gauge-invariant identity
\begin{equation}
    \sum_h F_{k}^{\mu \nu} F_{-k}^{\rho \sigma} = k^{\mu} k^{\rho} \eta ^{\nu \sigma}-k^{\nu} k^{\rho} \eta ^{\mu \sigma}-k^{\mu} k^{\sigma} \eta ^{\nu \rho}+k^{\nu} k^{\sigma} \eta ^{\mu \rho}\ .
\end{equation}
We find\footnote{Again, here we are ignoring zero-frequency gravitons, \textit{i.e.} $\omega_i \neq 0$ (or $k^0_i \neq 0$ in the rest frame of the massive particle). Then, the prescription for the external propagators is irrelevant.}
\begin{equation}
    \begin{split}
        d_{\mu \nu}\ell^\mu\ell^\nu & = 64 e^4 \frac{\ell\cdot F_{1}\cdot p  \left[ m^2 \ell\cdot F_{2}\cdot(p-k_1) + p\cdot k_1\, \ell\cdot F_{2} \cdot k_1 \right]}{(-2 p\cdot k_1)^2}\ , \quad \\
        d_{\mu} \ell^\mu & = - 32 e^4 \frac{\ell\cdot F_{1}\cdot p \ k_1\cdot F_{2}\cdot p}{(-2 p\cdot k_1)}\ ,\\
        c_{\mu} \ell^\mu & = - 16 e^4 \frac{p\cdot F_{1}\cdot F_{2} \cdot \ell}{(-2 p\cdot k_1)}\ ,\\
        c & = 16 e^4 \frac{p\cdot F_{1}\cdot F_{2} \cdot (p-k_1)}{(-2 p\cdot k_1)}\ , \\
        b & = d = 0\ .
    \end{split}
\end{equation}
The remaining coefficients are computed imposing Bose symmetry for the photons. Now, we can consider the classical limit at the level of the integrand. In the heavy-mass expansion, we have
\begin{equation}
    \begin{split}
        d_{\mu \nu} \ell^\mu \ell^\nu & = 64 e^4 \bM^2 \frac{\ell\cdot F_{1}\cdot \bu \ \ell\cdot F_{2}\cdot \bu}{(-2 \bu \cdot k_1)^2} +\cO(\bM^1)\ , \\
        c & = 16 e^4 \bM \frac{\bu \cdot F_{1}\cdot F_{2} \cdot \bu }{(-2 \bu \cdot k_1)} +\cO(\bM^0)\ , \\
    \end{split}
\end{equation}
and all the other coefficients contributing only to subleading order. Indeed, the Feynman integrals at leading order in the mass expansion look like
\begin{equation}
    \begin{split}
        j_{1,a_2,1,a_4}^{\mu_1 \dots \mu_n} &= \int\! \hd^D \ell\, \frac{\ell^{\mu_1}\, \cdots\, \ell^{\mu_n}}{[\ell^2]_+ [- 2 \bM\, \bu \cdot \ell]_+^{a_2+a_4} [- 2 \bM\, \bu \cdot (\ell+k_1)]_+} + \dots\\
        &= \frac{(-1)^{a_2+a_4}}{(- 2 \bM\, \bu \cdot k_1)^{a_2+a_4}}\int\! \hd^D \ell\, \frac{\ell^{\mu_1}\, \cdots\, \ell^{\mu_n}}{[\ell^2]_+ [- 2 \bM\, \bu \cdot (\ell+k_1)]_+} + \dots\ ,
    \end{split}
\end{equation}
where we used partial fractioning and we set to zero the resulting scaleless integrals.

Finally, we notice that the symmetrisation in $k_1\leftrightarrow k_2$ simply changes the overall sign of $c$, introduces an overall $(-1)^{a_2+a_4+1}$ in the integrals and modifies the $i \epsilon$ prescription of the massive propagator.
The integrals combine to localise the massive propagator on-shell, giving rise to the bubble integrals appearing in the first waveform computations from the heavy-mass approximation~\cite{Brandhuber:2023hhy,Herderschee:2023fxh}:
\begin{equation}
    \tilde{\cA}_4^1 (\bM\, \bu;k_1,k_2) = \frac{d_{\mu \nu} B^{\mu \nu}}{(- 2 \bM\, \bu \cdot k_1)^2} - \frac{c B}{(- 2 \bM\, \bu \cdot k_1)}\ ,
\end{equation}
where
\begin{equation}
    B^{\mu_1 \dots \mu_n} = - i \int\!\hd^D \ell\, \frac{\ell^{\mu_1}\, \cdots\, \ell^{\mu_n}}{[\ell^2]_+}\hdelta (- 2 \bM\, \bu \cdot (\ell+k_1))\ .
\end{equation}
We notice that performing the tensor reduction a là Passarino-Veltman the terms proportional to $u^\mu$ or $k_1^\mu$ are projected out by the external kinematics and the term proportional to $\eta^{\mu \nu}$ is the only giving a non-zero contribution. The result is proportional to the tree level.

\subsection{Generalised unitarity for the Fourier-loop integrand and tensor reduction}
\label{sec:Fourier_loop_integrand}

\begin{figure}[!t]
    \centering
    \includegraphicsbox[scale=1.5]{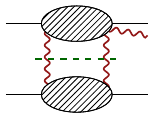} \>
    \includegraphicsbox[scale=1.5]{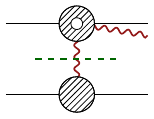} \>
    \caption{The singularities in the $q_i^2$ channels are probed by generalised unitarity cuts (in complex kinematics). At one loop we have contributions from the exchange of two gravitons (right) and one graviton (left), with the one-loop Compton amplitude appearing on one side of the cut. In electrodynamics (and, in general, in QED) there is no need for \textit{cut merging}, which is an analytic procedure to avoid double counting terms which contribute to both.}
    \label{fig:1loop_cuts}
\end{figure}

To avoid the appearance of spurious singularities in intermediate steps, we consider the loop integral together with the Fourier integral, and we study it as a genuine 2-loop integral with an exponential factor. In the previous two sections, we have computed the building blocks needed for the construction of the combined Fourier-loop integrand. This means that we supply the usual set of spanning cuts with additional cuts involving single-gravitons exchanges in the $q_{1,2}^2$ channels. The simplicity of the problem we are considering allows us to work with unitarity-like cuts, as shown in Figure~\ref{fig:1loop_cuts}, and avoid larger sets of \textit{spanning cuts} (for a review on generalised unitarity see \textit{e.g.} \cite{Bern:2011qt}).

\begin{figure}[!t]
    \centering
    \includegraphicsbox[scale=1.5]{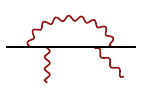} \> \qquad
    \includegraphicsbox[scale=1.5]{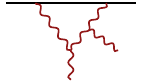} \>
    \caption{Two of the Feynman diagrams contributing to the one-loop gravitational Compton amplitude. The diagram on the left contributes to the QED amplitude as well, while the one on the right is characteristic of self-interacting bosons. The latter is the origin of the overlap of the poles and discontinuities in the $q_i^2$ channels.}
    \label{fig:1loop_compton}
\end{figure}

By inspecting the tree-level amplitudes (or in general the graphs structures and the derivative counting of the Feynman diagrams contributing to the process), we can easily put forward an ansatz for the one-loop waveform:\footnote{We introduced the short-hand notation $\int\! \hd^D q\, \hd^D \ell = \int_{\hat{q},\hat{\ell}}$.}
\begin{equation}
    \Delta \expval{\cW_h } = \frac{e^{i \omega\, n \cdot b_2}}{4\pi r} \!\int_{\hat{q},\hat{\ell}}\ \hdelta(2 \bp_1 \cdot q) \hdelta(2 \bp_2 \cdot (k-q))\, e^{i b \cdot q}\, \left(\cI_{\cW}^{D_1} + \cI^{R_1}_{\cW} +\cI_{\cW}^{D_2} + \cI^{R_2}_{\cW}\right)\ ,
\end{equation}
where
\begin{equation}
\label{eq:ansatz_discontinuity}
    \begin{split}
        \cI_{\cW}^{D_1} &= \sum_{p=0}^2 \sum_{m + n = 0}^3 D^{(1,p)}_{\mu_1 \dots \mu_m,\, \nu_{1}\dots \nu_n} \frac{\ell^{\mu_1}\cdots\, \ell^{\mu_m} q^{\nu_1}\cdots\, q^{\nu_n}}{\ell^2 (\ell-q)^2 (\bu_2 \cdot \ell)^{p}} (- i)\hdelta(\bu_1 \cdot \ell)\\
        &+ \sum_{p=0}^2 \sum_{m + n = 0}^3 D^{(2,p)}_{\mu_1 \dots \mu_m,\, \nu_{1}\dots \nu_n} \frac{\ell^{\mu_1}\cdots\, \ell^{\mu_m} q^{\nu_1}\cdots\, q^{\nu_n}}{\ell^2 (\ell-q+k)^2 (\bu_2 \cdot \ell)^{p}} (- i)\hdelta(\bu_1 \cdot \ell)+\dots
\end{split}
\end{equation} 
and
\begin{equation}
\label{eq:ansatz_pole}
    \begin{split}
        \cI_{\cW}^{R_1} = \sum_{m + n = 0}^3 R_{\mu_1 \dots \mu_m,\, \nu_{1}\dots \nu_n} \frac{\ell^{\mu_1}\cdots\, \ell^{\mu_m} q^{\nu_1}\cdots\, q^{\nu_n}}{q^2 (\ell-q)^2} (-i)\hdelta(\bu_1 \cdot \ell)+\dots\ .
    \end{split}
\end{equation}
The dots stand for terms that do not have a branch point and a pole in $q^2=0$ (the numerators cancel either $\ell^2$ or $(\ell -q)^2$, and $q^2$ in the denominator), respectively. This splitting makes sense only in electrodynamics: in gravity or, in general, theories with self-interacting bosons, we will have overlaps between the poles and the branch points in $q_{1,2}^2$. This becomes very clear from a quick inspection of the Feynman diagrams in the theory, as shown in Figure~\ref{fig:1loop_compton}. 
$\cI_{\cW}^{D_2}$ and $\cI_{\cW}^{R_2}$ are obtained by symmetry, $m_1 \leftrightarrow m_2$, $u_1 \leftrightarrow u_2$, $q\rightarrow k-q$. The computational task is further simplified by noticing that terms which have simultaneous singularities in $q_1^2$ and $q_2^2$ do not appear in QED (for example, at tree-level we do not have contributions of the form $\frac{1}{q^2 (k-q)^2}$ and at loop level we do not have pentagons appearing in the ansatz~\eqref{eq:ansatz_discontinuity}). Then, we can ignore these contributions throughout the computation and obtain the final result (after integration) by symmetrisation $m_1 \leftrightarrow m_2$, $u_1 \leftrightarrow u_2$ and $b_1 \leftrightarrow b_2$. Our results match the integrand of reference~\cite{Elkhidir:2023dco}.

Let us stress an interesting point. The reformulation of generalised unitarity by ansatz makes manifest the fact that the integrand is independent on the $i\epsilon$ prescription (\textit{i.e.} by the observable we are focusing on), as argued in reference~\cite{Caron-Huot:2023vxl}. Indeed, after writing down an ansatz for the integrand, we probe it by \textit{changing} the contour of integration to encircle a number of poles of the integrand, putting the corresponding particle on-shell in some complex kinematics -- irrespectively of the $i\epsilon$ prescription the original integral had. Perturbative unitarity (at the diagrammatic level) tells us that the corresponding integrand factorises.

The tensor structures appearing in the ansatz~\eqref{eq:ansatz_discontinuity}~and~\eqref{eq:ansatz_pole} are such that the $\ell$'s and $q$'s are contracted with either the field strengths or with external momenta, such that their scalar product does not appear in the denominators. If we choose to perform tensor reduction at the level of the loop integration only, we would find as an intermediate step a plethora of spurious poles complicating the Fourier integration (even numerically) \cite{Brandhuber:2023hhy,Herderschee:2023fxh,Bohnenblust:2023qmy}. In principle, the amplitude (the Fourier integrand, in general) can be rewritten in terms of a set of functions that are manifestly free of spurious poles on the physical sheet \cite{Bini:2024rsy}. On the other hand, in this work, we completely bypass this problem, by \textit{performing tensor reduction for $\ell$ and $q$ together}. Such reduction has been performed using the method proposed in reference~\cite{Anastasiou:2023koq}, which we briefly review in Appendix \ref{sec:A1}. For the sake of our computation, four external vectors appear, which are $ b, \bu_1, \bu_2, k$ and, in electrodynamics, we encounter tensors up to rank 4.
\section{IBP identities for Fourier integrals}
\label{sec:tensor}

In the previous section, we built the integrand of the waveform at one loop via generalized unitarity, and we reduced it to a linear combination of independent tensor structures, multiplied by scalar integrals. In this section, we reduce the number of integrals appearing to a minimal set of linearly independent terms -- the MIs -- to make the integration problem more accessible. Thus, we introduce integration-by-parts (IBPs) identities for the combined Fourier-loop integrals.

All the scalar integrals appearing belong to the integral family
    \begin{eqnarray}
\mathcal{I}_{a_1,a_2,a_3,a_4,a_5,a_6,a_7,a_8,a_9,a_{10},a_{11}}=\int_{\hat{q},\hat{\ell}} \ e^{ D_1} \prod_{i=1}^{11}\biggl(\frac{1}{D_i^{a_i}}\biggr) \ ,
\label{eq:topology}
\end{eqnarray}
where:
\begin{align}
   & D_1  = i q\cdot b, & & D_2 =   q^2, & &
    D_3  =   (q-k)^2 ,& & D_4 =  \bu_1 \cdot q, & & D_5 =  \bu_2 \cdot (k-q), &\nonumber\\  
   & D_6 =    \bu_1\cdot \ell  & & D_7 =  \bu_2\cdot \ell, & & D_8 =  \ell^2 ,& & D_9 =  (\ell+q)^2,& &  D_{10} =  (\ell+q- k )^2 , & & D_{11} = i  b \cdot \ell &
\end{align}
From the previous section, we know that $b^\mu$ can appear only in the numerator as it is produced by tensor reduction. Then, $D_{1,11}$ cannot appear as denominators and $a_{1,11}\leq 0$. Moreover, in the specific case of electrodynamics $D_{2,3}$ never appear together with $D_{9,10}$ as explained in Section~\ref{sec:Fourier_loop_integrand}. The on-shell measure forces $D_{4,5}$ to appear as delta functions. But, at the level of IBP relations, this is irrelevant as delta functions can be treated on the same footing as other propagators via reverse unitarity \cite{Anastasiou:2002yz,Anastasiou:2003gr,Herrmann:2021lqe}:
\begin{eqnarray}
    \frac{i}{(-1)^{s+1} s!}\hdelta^{(s)}(D_i) = \frac{1}{(D_i + i \epsilon)^{s+1}}- \frac{1}{(D_i - i \epsilon)^{s+1}} \ . 
\end{eqnarray}
Finally, the classical limit localises one matter line per loop (this is manifest in the heavy-mass approach \cite{Brandhuber:2021eyq} and in the worldline approach \cite{Kalin:2020mvi,Mogull:2020sak}), and we find that either $D_6$ or $D_7$ are cut. Hence, the scalar integrals can be divided into four families:
\begin{eqnarray}
\label{eq:QED_sector}
\mathcal{I}_{a_1,a_2,a_3,1,1,1,a_7,a_8,a_9,a_{10},a_{11}}^{\,\bu_1}& =&  \int_{\hat{q},\hat{\ell}}\,  e^{ D_1} \frac{D_1^{-a_1}D_2^{-a_2}D_3^{-a_3}D_{11}^{-a_{11}}\hdelta(D_4)\hdelta(D_5)\hdelta(D_6)}{D_7^{a_7}D_8^{a_8}D_9^{a_9}D_{10}^{a_{10}}}\ ,  \\
\mathcal{I}_{a_1,a_2,a_3,1,1,1,a_7,a_8,a_9,a_{10},a_{11}}^{\,\bu_1, C}& =&  \int_{\hat{q},\hat{\ell}}\,  e^{ D_1} \frac{D_1^{-a_1}D_2^{-a_2}D_7^{-a_7}D_8^{-a_8}D_{10}^{-a_{10}}D_{11}^{-a_{11}}\hdelta(D_4)\hdelta(D_5)\hdelta(D_6)}{D_3^{a_3}D_9^{a_9}}\ ,\nonumber \\ \label{eq:QED_sector_2}
\end{eqnarray}
$\mathcal{I}^{\,\bu_2}$ that can be obtained from $\mathcal{I}^{\, \bu_1}$ by exchanging $D_6 \leftrightarrow D_7$, and $\mathcal{I}^{\,\bu_2,C}$ that can be obtained from $\mathcal{I}^{\, \bu_1, C}$ by exchanging $D_6 \leftrightarrow D_7$, $D_2 \leftrightarrow D_3$, $D_9 \leftrightarrow D_{10}$.\footnote{Here, we are using a standard notation for which the scalar products appearing in the numerators have always $a_i\leq 0$ in the scattering amplitude.}

In dimensional regularisation, IBPs can be performed by imposing that the total derivative w.r.t. the Fourier or the loop momenta vanishes under the integral sign:
\begin{equation}
    \int_{\hat{q},\hat{\ell}} \frac{\partial}{\partial{\{\ell^\mu ,q^\mu\}}} \biggl( \ e^{ D_1} \frac{v^\mu}{\prod_{i=1}^{11}D_i^{a_i}}\biggr) =  0 
\end{equation}
where $v^\mu$ can be a linear combination of loop momenta and external momenta (a similar approach has been adopted in \cite{Luo:2019szz}). The generation of IBP identities in the presence of exponential functions can be performed algorithmically from the IBP relations generated from the integral family where the exponential is not present. Expanding the total derivative under the integral sign, and recasting the various terms, we have:\footnote{Notice that if we take the derivative w.r.t. $\ell$ the second and the third terms vanish and find the usual IBPs relations.} 
\begin{equation}   
 \int_{\hat{q},\hat{\ell}} e^{ D_1}\biggl[   \frac{\partial}{\partial q^\mu}\biggl(\frac{v^\mu}{\prod_{i=1}^{11}D_i^{a_i}}\biggr)+\frac{\partial}{ \  \partial{q^\mu}}\biggl( D_1\frac{v^\mu}{\prod_{i=1}^{11}D_i^{a_i}}\biggr)
 - D_1 \frac{\partial}{ \  \partial q^\mu}\biggl( \frac{v^\mu}{\prod_{i=1}^{11}D_i^{a_i}}\biggr)  \biggr] = 0  \ . 
 \label{eq:ibp_1l}
\end{equation}
Equation \eqref{eq:ibp_1l} can be rewritten in terms of standard (two-loop) IBP relations -- denoted as \text{IBP}$[a_1,\ldots,a_{11}]$ -- of an integral family containing the same set of denominators but not exponentials:
\begin{eqnarray}
    \text{IBP}[a_1,\ldots, a_{11}]+ \text{IBP}[a_1-1,\ldots, a_{11}]-D_1 \text{IBP}[a_1,\ldots, a_{11}] = 0 \ , 
\end{eqnarray}
where we generate $\text{IBP}[a_1,\ldots, a_{11}]$ using \texttt{LiteRed}~\cite{Lee:2012cn}.

In principle, one can generate and solve the IBP relations in all the sectors appearing.\footnote{An integral sector  $\mathcal{S}=(\sigma_1,\ldots \sigma_{11})$, where $\sigma_{i} = 0,1$, is the set of points $(a_1,\ldots, a_{11})$ in $\mathbb{Z}^{11}$ such that  $\Theta (a_i-1/2) = \sigma_i$. In particular, all the integrals of a given sector have the same set of denominators.} However, it is useful to restrict our attention to the subsets where we expect to find MIs, called \textit{non-zero sectors}. These can be isolated by counting the number of MIs associated with each sector, as pointed out in references~\cite{Lee:2013hzt,Bitoun:2017nre}. This counting is most efficiently carried out in Baikov representation \cite{Baikov:1996iu,Frellesvig:2017aai}, in which equation~\eqref{eq:topology} is written as a parametric integral over the denominators $z_i = D_i$:
\begin{equation}
    I_{a_1,\ldots a_{11}} \propto \int\! u(\mathbf{z})   \prod_{i=1}^{11}\biggl(\frac{dz_i}{z_i^{a_i}}\biggr) \ , \qquad u(\mathbf{z}) = \ e^{z_{11}} B(\mathbf{z})^\gamma \ , 
\end{equation} 
where $B(\mathbf{z})=\det\mathbf{G}$ is the Gram determinant of the loop momenta and the external vectors ($q,\ell,u_1,u_2,k,b$) and $\gamma = (D-7)/2$. The number of MIs of a given sector $\mathcal{S}$ is obtained as the number of critical points of the regulated function $u_\mathcal{S} = u(\mathbf{z}) \prod_{i=1}^{11} z_i^{\sigma_i \rho_i}$~\cite{Mastrolia:2018uzb,Frellesvig:2019uqt}, where the $\rho_i$ are analytic regulators, and it is given as the number of zeroes of\footnote{In practice, we can fix the $\rho_i$'s to non-integer numbers.}
\begin{eqnarray}
\omega_{\mathcal{S}} = \dd\log (u_\mathcal{S}) = \sum_{i=1}^{11} \partial_{z_i}(u_\mathcal{S}) \dd z_i = 0  \ .
\end{eqnarray}
This computation in practice can be done numerically in \texttt{Mathematica} by solving the system of equations w.r.t. $z_i$, or using the Julia package \texttt{HomotopyContinuation}~\cite{HomotopyContinuation.jl}.

Then, restricting to the non-zero sectors, we solved the IBP system twice, using \texttt{LiteRed}~\cite{Lee:2012cn}, and with an in-house routine on \texttt{FiniteFlow}~\cite{Peraro:2019svx} using Laporta method \cite{Laporta:2000dsw}, finding agreement between the two results. The decomposition in terms of MIs has been independently checked using intersection numbers \cite{Mastrolia:2018uzb}, using the method developed in \cite{Brunello:2023rpq}.

Focusing on the relevant sector in equation~\eqref{eq:QED_sector} (with $a_{1,2,3,11}\leq 0$), and in ~\eqref{eq:QED_sector_2} (with $a_{1,3,7,8,10,11}\leq 0$), we find respectively 16 and 2 MIs, but only a subset of them enters the electrodynamics calculation. Moreover, we notice that integrals with different powers of $D_1$ in the numerators are related by a differentiation w.r.t. $b^\mu$: 
\begin{equation}
\mathcal{I}^{\, \bu_i}_{-a_1, a_2,\ldots, a_{10},0} = \delta_b^{(a_1)} \mathcal{I}^{\, \bu_i}_{0, a_2,\ldots, a_{10},0 }\ ,
\end{equation}
where
\begin{equation}
    \qquad \delta_b^{(n)}  = b^{\mu_1} \cdots b^{\mu_n} \frac{\partial}{\partial b^{\mu_1}}\cdots\frac{\partial}{\partial b^{\mu_n}} \ ,
\end{equation}
and we showed for simplicity only the case $a_{11} = 0$, which is relevant for our computation.

The one-loop waveform in electrodynamics can be decomposed in terms of ten MIs of the first family:
\begin{equation}
\begin{split}
    \mathcal{J}^{\, \bu_1}_{1+n} & = \mathcal{I}^{\, \bu_1}_{-n,0,0,1,1,1,0,1,0,1,0} = \delta_b^{(n)} \, \mathcal{F}\left[   \includegraphicsbox[scale=.7]{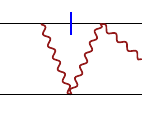} \right]\ ,\\
    \mathcal{J}^{\, \bu_1}_{5+m} & = \mathcal{I}^{\, \bu_1}_{-m,0,0,1,1,1,1,1,0,1,0} = \delta_b^{(m)} \, \mathcal{F}\left[   \includegraphicsbox[scale=.7]{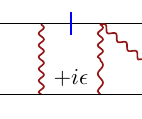} \right]\ ,\\
    \mathcal{J}^{\, \bu_1}_{7+m} & = \mathcal{I}^{\, \bu_1}_{-m,0,0,1,1,1,0,1,1,0,0} = \delta_b^{(m)} \, \mathcal{F}\left[ \includegraphicsbox[scale=.7]{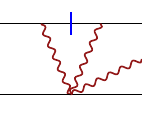} \right]\ ,\\
    \mathcal{J}^{\, \bu_1}_{9+m} & = \mathcal{I}^{\, \bu_1}_{-m,0,0,1,1,1,1,1,1,0,0} = \delta_b^{(m)} \, \mathcal{F} \left[\includegraphicsbox[scale=.7]{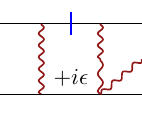} \right]\ .
\end{split}
\end{equation}
and two MIs of the second one:
\begin{equation}
\begin{split}
    \mathcal{J}^{\, \bu_1, C}_{1+m} & = \mathcal{I}^{\, \bu_1}_{-m,0,1,1,1,1,0,0,1,0,0} = \delta_b^{(n)} \, \mathcal{F}\left[ \includegraphicsbox[scale=.7]{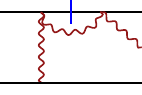} \right]\ ,
    \end{split}
\end{equation}
with $n=0,\dots 3$, $m=0,1$ and $\mathcal{F}$
is a short-hand notation for the FT to impact parameter space:
\begin{eqnarray}
    \mathcal{F}[f(q)] = \int_{\hat{q}}  e^{i q \cdot b}\,  \hat{\delta }(\bu_1\cdot q) \, \hat{\delta }(\bu_2\cdot (q-k))\, f(q) \ . 
\end{eqnarray}
Hence, it is sufficient to compute the FT of $\mathcal{J}_{1,5,7,9}^{\, \bu_1}$, and of $\mathcal{J}_{1}^{\, \bu_1, C}$. The final result for the waveform can be written as
\begin{equation}
    \Delta \expval{\cW_h }(\omega,\vec{n}) = \frac{e^{i \omega\, n \cdot b_2}}{4\pi r}  \sum_{i = 1}^{10} \left(c^{\bu_1}_i \mathcal{J}^{\, \bu_1}_{i} + c^{\bu_2}_i \mathcal{J}^{\, \bu_2}_{i}\ \right)+\frac{e^{i \omega\, n \cdot b_2}}{4\pi r}  \sum_{i = 1}^{2} \left(c^{\bu_1, C}_i \mathcal{J}^{\, \bu_1, C}_{i} + c^{\bu_2, C}_i \mathcal{J}^{\, \bu_2, C}_{i}\ \right),
\end{equation}
where the coefficients $c^{\bu_1}_i$ can be found in the ancillary file \texttt{waveformED.m} and $c^{\bu_2}_i$ are obtained by symmetry.

\subsection{Evaluating the Fourier integrals}
In Appendix~\ref{sec:A2}, we gave the technical details for the evaluation of all the one-loop integrals entering a waveform computation. The relevant one-loop integrals, with retarded (matter) propagators, for the case of electrodynamics are:\footnote{We remind the reader that in comparing these results with the integrals in references~\cite{Brandhuber:2023hhy,Herderschee:2023fxh}, one needs to sum both the integrals computed with principal value prescriptions and those with two on-shell matter propagators, which have been presented in the latest versions of these papers.}
\begin{equation}
\begin{split}
\includegraphicsbox[scale=.7]{figures/compton_mis_v2} 
&  = -\frac{i w_1}{4 \pi (-q_2^2) }+\cO\left(\epsilon^1\right) , \\
\includegraphicsbox[scale=.7]{figures/triu1}
&  = \frac{ \pi + 2 i  \log \frac{w_1+\sqrt{w_1^2-q_2^2}}{\sqrt{-q_2^2}}}{8 \pi 
   \sqrt{w_1^2-q_2^2}}+\cO\left(\epsilon^1\right),\\
\includegraphicsbox[scale=.7]{figures/triu1bl}&  = \frac{1}{8 \sqrt{(-q_1^2)}}+\cO\left(\epsilon^1\right),\\
\includegraphicsbox[scale=.7]{figures/boxu1}
&  = 
\frac{i}{8 \pi (-q_2^2)\sqrt{\gamma^2-1}} \left(\frac{-q_2^2}{w_1 \mu_{\rm IR}}\right)^{-2\epsilon} \left[\frac{1}{\tilde{\epsilon}}-\log \left(\gamma^2-1\right) -2\log \left(\gamma+\sqrt{\gamma^2-1}\right)+i \pi \right]+\cO\left(\epsilon^1\right)\\
\includegraphicsbox[scale=.7]{figures/boxu1bl}&  =
\frac{i}{8 \pi (-q_1^2)\sqrt{\gamma^2-1}} \left(\frac{-q_1^2}{w_2 \mu_{\rm IR}}\right)^{-2\epsilon} \left[\frac{1}{\tilde{\epsilon}}-\log \left(\gamma^2-1\right) +i \pi \right]+\cO\left(\epsilon^1\right), \\
\end{split}
\label{eq:neededints}
\end{equation}
where, we defined $\tilde{\epsilon} =  \epsilon\,e^{(\gamma_E - \log 4\pi)\epsilon}$, $\gamma_E$ is the Euler-Mascheroni constant and $\mu_{\rm IR}$ is the infrared scale introduced in dimensional regularisation.
Four of the five integrals appearing in equation \eqref{eq:neededints} can be recast (at least up to corrections of order $\epsilon^1$) as\footnote{The Fourier integrals $\mathcal{F}\left[(-(k-q)^2)^\alpha\right]$ can be obtained performing a trivial shift on the integrated momentum mismatch, effectively changing $w_2 \to w_1$. This also gives an additional phase $e^{i \omega n \cdot b}$, which combines with the overall phase in equation~\eqref{eq:waveform_frequency_space} to give $e^{i \omega n \cdot b_1}$.} 
\begin{equation}
    \mathcal{F}\left[(-q^2)^\alpha\right] = \frac{\Omega_{D-4}}{\sqrt{\gamma^2-1} (2\pi)^{D-2}} \int_{\mathcal{C}}\!\dd^2 z\, z_v^{D-4} e^{ - i z_b \sqrt{-b^2}}\, \left(z_v^2+z_b^2+\frac{w_2^2}{\gamma^2-1}\right)^{\alpha}\ ,
\end{equation}
where $\Omega_{D-4}= \frac{2\pi^{\frac{D-3}{2}}}{\Gamma\left[\frac{D-3}{2}\right]}$ is the ``volume'' of the $(D-4)$-sphere and $\mathcal{C}$ is the integration contour mentioned above, \textit{i.e.} $(z_v,z_b) \in \mathbb{R}_+\times \mathbb{R}$. For the boxes $\mathcal{J}_{5}^{\, \bu_2}$ and $\mathcal{J}_{9}^{\, \bu_1}$, we have $\alpha = -1-2 \epsilon$ -- in this case the result is valid only at $\cO(\epsilon^1)$ -- while for the triangle $\mathcal{J}_{7}^{\, \bu_1}$ we have $\alpha = \frac{D-5}{2}$. The integral is computed in terms of Bessel-$K$ functions:
\begin{equation}
\label{eq:FT_q_power}
    \mathcal{F}\left[(-q^2)^\alpha\right] = -\frac{2}{\sqrt{\gamma^2-1}\, (4\pi)^{\frac{D}{2}-1}\, \Gamma (-\alpha )} \left(\frac{2 z}{-b^2}\right)^{\alpha +\frac{D}{2}-1} K_{\alpha+\frac{D}{2} -1}\left(z\right)\ ,
\end{equation}
where we introduced the variable
\begin{equation}
    z = \frac{w_2 \sqrt{-b^2}}{\sqrt{\gamma^2-1}}\ ,
\end{equation}
which scales as $\sim p_\infty^0$ in the small velocity limit, where $p_\infty = \sqrt{\gamma^2-1}$ is the  expansion parameter~\cite{Bini:2023fiz}. The other triangle is more subtle. It can be expanded in the small velocity limit $p_\infty\sim 0$, making each term in the series of the form~\eqref{eq:FT_q_power}. Then, we find
\begin{equation}
    \begin{split}
        \mathcal{J}_{1} & = \frac{i z}{2(4\pi)^{\frac{3}{2}}\sqrt{-b^2\, p_\infty}} \sum_{n=0}^\infty \left(-\frac{z p_\infty}{2}\right)^n \left[\frac{K_n \left(z\right)}{\Gamma\left[n+\frac{3}{2}\right]}-i \, \frac{K_{n-\frac{1}{2}} \left(z\right)}{\sqrt{\frac{z p_\infty}{2}}\, n!}\right]\\
        & = \frac{i}{16 \pi \sqrt{-b^2\, p_\infty}} \left\{z \!\int_0^{\infty}\! \dd x \left[ e^{- z \cosh x}\, \mathbf{H}_{-1} \left(z \sqrt{p_\infty} \sinh x\right)\right] - i \, \frac{e^{-z \sqrt{1+p_\infty}}}{\sqrt{p_\infty}}\right\}\ ,
    \end{split}
\end{equation}
where $\mathbf{H}_{\alpha}(z)$ is the Struve function with index $\alpha$. Notice that the integral representation can be efficiently evaluated numerically, as it converges very fast.
\section{Conclusions}
\label{sec:conclusions}

In this paper, we presented a direct strategy to compute scattering waveforms derived from the KMOC formalism. We clarified the heavy-mass expansion of the tree-level amplitudes, showing that non-trivial contributions containing delta functions can contribute to the classical observables.
We exploited the analytic structure of the in-in correlator to single out long-range contributions to the waveform. Moreover, we by-passed the appearance of spurious poles, by performing tensor reduction for combined Fourier-loop tensor structures. Finally, we introduced IBP relations to reduce the number of master integrals, treating the momentum mismatches on the same footing as loop momenta. This strategy allowed us to compute the one-loop waveform in \textit{electrodynamics} and paves the way for the analytic evaluation of gravitational waveform for arbitrary velocities (within the D'Eath bound \cite{DEath:1976bbo,DiVecchia:2023frv}), which will be presented elsewhere \cite{Brunello:toappear}.

Our investigation opens the way for a few interesting questions. In the heavy-mass expansion, the structures involving principal-value propagators have been understood in all cases with a single massive source and any number of gluons (or photons, through decoupling identities, or gravitons, through double-copy \cite{Bern:2008qj}) \cite{Brandhuber:2021kpo,Brandhuber:2021bsf,Brandhuber:2022enp}. A systematic understanding of classical terms involving delta function contributions is still lacking. Moreover, the computation of the Fourier integrals is performed by computing the loop integrals first (with differential equations or by direct evaluation). The differential equations for the combined Fourier and loop integrals involve integration kernels with Bessel functions with integer and half-integer indices, most probably giving iterated Bessel functions in the $\epsilon$ expansion (for example, see \cite{Brunello:2023fef}). At tree level, performing the FT to the time domain simplifies the class of functions appearing \cite{Jakobsen:2021lvp} (see also \cite{Herderschee:2023fxh} for a detailed discussion). It will be interesting to investigate whether such simplification holds at loop level as well and if these integrals can be treated on the same footing as ordinary $(L+1)$-loop integrals.\footnote{We would like to thank Aidan Herderschee, Fei Teng and Radu Roiban for private communications on this point.} Finally, throughout this paper, we have selected terms in the amplitude which have branch points at $q_i^2$, but we have not computed directly the discontinuity along their branch cuts.

\acknowledgments

We thank Gang Chen, Manoj K. Mandal, Pierpaolo Mastrolia, Donal O'Connell, Matteo Sergola for discussions. SDA would like to thank Aidan Herderschee, Radu Roiban, Fei Teng for discussions on related projects. GB would like to thank Giulio Crisanti, Mathieu Giroux, Pierpaolo Mastrolia, Manoj K. Mandal, Sid Smith for discussions on related projects. In particular, we would like to thank David A. Kosower for several inputs along the finalisation of this project. We also thank Donato Bini, Thibault Damour, Asaad Elkhidir, Harald Ita, Pierpaolo Mastrolia, Matteo Sergola, Donal O'Connell, Fei Teng, Sid Smith for insightful comments on the draft and spotting typos.
Our research is supported by the European Research Council, under grant ERC–AdG–88541.

\appendix

\section{Passarino-Veltman reduction}
\label{sec:A1}
In this appendix, we introduce briefly the Passarino-Veltam reduction a là Anastasiou, Karlen and Vicini \cite{Anastasiou:2023koq}. The $D$-dimensional space can be decomposed in a \textit{physical subspace} spanned by a set of $E$ linearly-independent external vectors, $p_1,\dots, p_E$, and an the orthogonal subspace of dimension $D_\perp= D-E$. We can write down projectors onto these two subspaces. In particular, the projector onto the physical space is given by
\begin{equation}
    \eta_{\parallel}^{\mu\nu} = \sum_{i=1}^E p_i^\mu \langle p_i^\nu \rangle 
\end{equation}
where $\langle p_i^\nu \rangle $ are called \textit{dual vectors} and are defined such that $p_i^\mu \langle p_{j,\mu} \rangle = \delta_{ij}$. A closed form of these vectors is given in term of the Gram matrix:
\begin{eqnarray}
     \langle p_i^\mu \rangle  = \mathbf{G}^{-1}_{ij}\  p_j^\mu
\end{eqnarray}
where $\mathbf{G}_{i j} = p_i\cdot p_j$ and $i,j=1,\dots E$.
Trivially, the projector onto the orthogonal space is
\begin{eqnarray}
    \eta_{\perp}^{\mu\nu} = \eta^{\mu\nu}-\eta_\parallel^{\mu\nu} \ .
\end{eqnarray}
For a given tensor integral of rank $R$, $I^{\mu_1\ldots \mu_R}$, its tensor decomposition can be written in closed form as:
\begin{eqnarray}
    I^{\mu_1,\ldots, \mu_R}= I_{\nu_1,\ldots, \nu_R} \mathcal{T}\biggl\{ \prod_{i=1}^{R}\eta_{\parallel}^{\nu_i \mu_i} \biggr\} \ ,
\end{eqnarray}
where $\mathcal{T}$ is an ordering symbol, resembling the Wick contractions, defined in reference~\cite{Anastasiou:2023koq} as:
\begin{equation}
\begin{split}
    \mathcal{T}\biggl\{\prod_{i=1}^{R}\eta_{\parallel}^{\nu_i \mu_i}\biggr\} & = 
    \ \ \ \prod_{i=1}^{R}\eta_{\parallel}^{\nu_i \mu_i} \\
    & 
    + \ \wick{\c \eta_{\parallel}^{\nu_1 \mu_1}\c \eta_{\parallel}^{\nu_2 \mu_2}}\prod_{i=3}^{R}\eta_{\parallel}^{\nu_i \mu_i} + \text{permutations} \\ 
    & 
    +\  \wick{\c1 \eta_{\parallel}^{\nu_1 \mu_1}\c1 \eta_{\parallel}^{\nu_2 \mu_2}\c2 \eta_{\parallel}^{\nu_3 \mu_3}\c2 \eta_{\parallel}^{\nu_4 \mu_4}}\prod_{i=5}^{R}\eta_{\parallel}^{\alpha_i \mu_i}+ \text{permutations} +\dots \ ,
\end{split}
\end{equation}
where the contractions are defined as
\begin{eqnarray}
     \prod_{i=1}^{n/2} \wick{\c \eta_{\parallel}^{\nu_i \mu_i}\c \eta_{\parallel}^{\nu_{i+1} \mu_{i+1}}} = \prod_{i=1}^{n/2} \eta_\perp^{\nu_i \nu_{i+1} }\bigl\langle  \prod_{i=1}^{n/2} \eta_\perp^{\mu_i \mu_{i+1} } \bigr\rangle \ .
\end{eqnarray}
Here, $\langle  \prod_{i=1}^{n/2} \eta_\perp^{\mu_i \mu_{i+1} } \bigr\rangle$ is the dual of the product of orthogonal projectors, \textit{i.e.} when contracted with the same number of orthogonal projectors give 1 or 0 if the indices are distributed in the same way or differently, respectively.

We will apply this tensor reduction to the loop momenta and Fourier momentum mismatches. The physical subspace is spanned by the four vectors $b^\mu$, $\bu_1^\mu$, $\bu_2^\mu$ and $k^\mu$. In electrodynamics we encounter tensors reduction up to rank 4.
\section{The pentagon integrals from differential equations}
\label{sec:A2}
\begin{figure}[h]
    \centering
    \includegraphicsbox[scale=1]{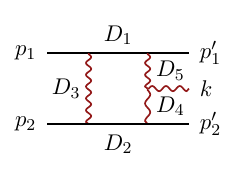}
     \caption{Topology for the one-loop 5-points integrals} \label{fig:topo1L}
\end{figure}
The goal of this section is to compute the pentagon integrals appearing in the one-loop amplitude in the heavy-mass limit via differential equations.

All the integrals appearing are related by partial fractioning to the integral family shown in Figure~\ref{fig:topo1L}:
\begin{equation}
    \mathcal{I}_{a1,a2,a3,a4,a5}= e^{\gamma_E \epsilon}\int_\ell \frac{1}{D_1^{a_1}D_2^{a_2}D_3^{a_3}D_4^{a_4}D_5^{a_5}}
    \end{equation}
where:
\begin{align}
    D_1 =  \bu_1\cdot \ell\ ,&  & D_2  = \bu_2 \cdot \ell  ,& &  D_3  = \ell^2, & & D_4 = (\ell-q_2)^2 ,& & D_5= (\ell+q_1)^2  \ . 
\end{align}
In particular, in the heavy-mass expansion, the relevant contributions to the classical limit appear always with either $D_1$ or $D_2$ localised by a delta function:
\begin{align}
    \mathcal{I}^{\, \bu_1}_{1,a_2,a_3,a_4,a_5}= e^{\gamma_E \epsilon}\int_\ell \frac{\hdelta(D_1)}{D_2^{a_2}D_3^{a_3}D_4^{a_4}D_5^{a_5}}, & &  \mathcal{I}^{\, \bu_2}_{a_1,a_2,a_3,a_4,a_5}=e^{\gamma_E \epsilon}\int_\ell \frac{\hdelta(D_2)}{D_1^{a_1}D_3^{a_3}D_4^{a_4}D_5^{a_5}}
\end{align}
By solving the associated sets of IBPs with \texttt{LiteRed}\cite{Lee:2012cn}, we find that each of the two families has 10 MIs and, since the other can be obtained by symmetries, we can focus on one of them only.
We are interested in finding a solution to the associated system of differential equations. It is useful to find a pure basis of integrals \cite{Henn:2013pwa}, which is a basis in which the $\epsilon$-dependence of the DEs is factorised, and at each order in $\epsilon$ our integrals appear as a sum of uniform-transcendentality functions.
A quantity that plays a crucial role in finding a pure basis is the leading singularity (LS) of an integrand, which can be found by replacing the integration paths with contour integrals around the poles of the integrand.
At one loop, an integral with $n$ or $n-1$ legs is pure if it has unit LS in $n$-dimensions \cite{Spradlin:2011wp, Flieger:2022xyq}. Moreover, integrals in different space-time dimensions can be related via dimensional-shift identities if their dimension differs by multiples of two \cite{Tarasov:1996br,Lee:2009dh}. Hence, we can find a pure basis by rescaling each $n$-point integral by its LS, computed around $d_0=n$ (or $n-1$) dimensions, and then relating them to their 4-dimensional representation via dimensional-shift identities. In particular, bubble integrals are pure in $d_0=2$, triangles and boxes around $d_0=4$, and pentagons around $d_0=6$. The LS can be easily computed in Baikov representation, by looking at the maximal cut for each integral \cite{Dlapa:2021qsl}.
Putting everything together we find:
\begin{align}
&\mathcal{G}_1 
\ = \ 
\frac{\epsilon (1-2 \epsilon)}{w_1}\ \mathcal{I}^{\, \bu_1}_{1,0,0,1,0} \ , &   &\mathcal{G}_2 \ =\ \epsilon^2 \sqrt{-q_1^2} \ \mathcal{I}^{\, \bu_1}_{1,0,1,0,1} & \nonumber\\
&\mathcal{G}_3 
\ = \ 
\epsilon^2 \sqrt{w_1^2-q_2^2}\ \mathcal{I}^{\, \bu_1}_{1,0,1,1,0} \ , & &
  \mathcal{G}_4 \ =\ \epsilon^2 \sqrt{\gam^2-1}\  \mathcal{I}^{\, \bu_1}_{1,1,0,0,1}  &\nonumber\\
&\mathcal{G}_5 
\ = \ 
\epsilon^2 \sqrt{\gam^2-1}\ \mathcal{I}^{\, \bu_1}_{1,1,0,1,0} \ , & &
  \mathcal{G}_6 \ =\ \epsilon^2 q_1^2 w_1\ \mathcal{I}^{\, \bu_1}_{1,0,1,1,1}  &\nonumber\\
&\mathcal{G}_7 
\ = \ 
\epsilon^2 w_1 w_2 \ \mathcal{I}^{\, \bu_1}_{1,1,0,1,1} \ , &
&  \mathcal{G}_8 \ =\ \epsilon^2 q_1^2 \sqrt{\gam^2-1}\ \mathcal{I}^{\, \bu_1}_{1,1,1,0,1}  &\nonumber\\
& \mathcal{G}_9 
\ = \ 
\epsilon^2 q_2^2 \sqrt{\gam^2-1}\ \mathcal{I}^{\, \bu_1}_{1,1,1,1,0} \ , 
& &  \mathcal{G}_{10} \ =\ \epsilon^2 \sqrt{\det\mathbf{G}} \ \mathcal{I}^{\, \bu_1, 6-2\epsilon}_{1,1,1,1,1} & 
\end{align}
where $\det\mathbf{G}$ is the Gram determinant of the external momenta $\bu_{1,2}$ and $q_{1,2}$. \\
This basis matches the one obtained in \cite{Bohnenblust:2023qmy}. Then, one gets a system of DEs in canonical form, which can be combined  in a total differential: 
\begin{equation}
    d \mathcal{G} = \epsilon d\mathbb{A} \mathcal{G} \ , \qquad d \mathbb{A}= \hat{\mathbb{A}}_{w_1} dw_1+ \hat{\mathbb{A}}_{w_2} dw_2+ \hat{\mathbb{A}}_{-q_1^2} d(-q_1^2)+ \hat{\mathbb{A}}_{-q_2^2} d(-q_2^2) + \hat{\mathbb{A}}_{\gam} d\gam \ , 
\end{equation} 
where the differential matrix is in $\dd \log$ form, and can be written as:
\begin{equation}
    d \mathbb{A} = \sum_{i=1}^{42} \mathbb{M}_i d\log (\eta_i) \ . 
\end{equation} 
where: $\eta_i$ are called letters, and they contain all the kinematic dependence of the DEs, and $ \mathbb{M}_i$ are constant matrices.
The alphabet consists of 42 letters that can be found using the method developed in \cite{Jiang:2024eaj}. There are 16 rational letters:
\begin{align}
& \eta_1 = -q_1^2 , 
& & 
\eta_2 = -q_2^2 , 
& & 
\eta_3 = \dom_1 , 
& & 
\eta_4 = \dom_2 
& \nonumber \\
& \eta_5 = \gam+1 , 
& & 
\eta_6 = \gam-1 , 
& & 
\eta_7= q_1^2 - q_2^2, 
& & 
\eta_8 = \dom_1^2-q_2^2 ,
\nonumber \\
& \eta_9 = \dom_2^2 -q_1^2,
& & 
\eta_{10} =-\dom_1^2-\dom_2^2+2 \dom_1 \dom_2 \gam,
& & 
\eta_{11} = -\dom_2^2-q_1^2 (\gam^2-1),
& & 
\eta_{12} = -\dom_1^2-q_2^2 (\gam^2-1) \nonumber 
\end{align}
\vspace{-1 cm}
\begin{align}
& \eta_{13} = -4 q_1^2 \dom_1^2 - (q_1^2-q_2^2)^2 , 
& \nonumber \\
&
\eta_{14} = -4 q_2^2 \dom_2^2 - (q_1^2-q_2^2)^2 , 
& \nonumber \\ 
& \eta_{15} = 2 q_1^2 q_2^2 \dom_1 \dom_2 \gam - q_1^4 \dom_1^2 - q_2^4 \dom_2^2 , 
& \nonumber \\
&\eta_{16} = (q_1^2 - q_2^2)^2 (\gam^2-1)+ 4(q_1^2 \dom_1 (\dom_2 \gam -\dom_1)+q_2^2 \dom_2 (\dom_1\gam - \dom_2)+\dom_1^2 \dom_2^2 ), \ \ \ \ \ \ \ \ \ \ \  \  \ \ \ \qquad \ \ 
\end{align}
and 26 algebraic letters:
\begin{align}
    & \eta_{17} 
    = \dom_2+\sqrt{\eta_9}&
    & \eta_{18} 
    = \dom_1+\sqrt{\eta_8} & \nonumber \\
    & \eta_{19} 
    = \gam+\sqrt{\eta_5 \eta_6} &
    & \eta_{20} 
    =- \frac{\dom_2 -\dom_1 \gam -\dom_1\sqrt{\eta_5 \eta_6}}{\dom_2 -\dom_1 \gam +\dom_1\sqrt{\eta_5 \eta_6}}& \nonumber \\
    & 
    \eta_{21} 
    =- \frac{q_1^2 \dom_1 -q_2^2 \dom_2 (\gam - \sqrt{\eta_5 \eta_6})}{q_1^2 \dom_1 -q_2^2 \dom_2 (\gam + \sqrt{\eta_5 \eta_6})} &
     & \eta_{22} 
    =- \frac{q_1^4 - q_1^2 q_2^2+ 2 q_2^2 \dom_2( \dom_2-\sqrt{\eta_9}) }{+q_1^4 - q_1^2 q_2^2+ 2 q_2^2 \dom_2( \dom_2+\sqrt{\eta_9}) }&\nonumber \\
    & 
    \eta_{23} 
    =-\frac{q_2^4 - q_1^2 q_2^2 -2 q_1^2\dom_1 (\dom_1-\sqrt{\eta_8}))}{q_2^4 - q_1^2 q_2^2 -2q_1^2 \dom_1 (\dom_1+\sqrt{\eta_8})} &
     & 
    \eta_{24} 
    =-\frac{\dom_1-\sqrt{\eta_2 \eta_5 \eta_6}}{\dom_1+\sqrt{\eta_2 \eta_5 \eta_6}}&\nonumber \\
     & 
    \eta_{25} 
    =-\frac{\dom_2-\sqrt{\eta_1 \eta_5 \eta_6}}{\dom_2+\sqrt{\eta_1 \eta_5 \eta_6}}&
     & 
    \eta_{26} 
    =-\frac{\dom_1\gam -\sqrt{\eta_5 \eta_6 \eta_8}}{\dom_1\gam +\sqrt{\eta_5 \eta_6 \eta_8}}&\nonumber \\
     & 
    \eta_{27} 
    =-\frac{\dom_2 \gam -\sqrt{\eta_5 \eta_6 \eta_9}}{\dom_2 \gam +\sqrt{\eta_5 \eta_6 \eta_9}}&
    & 
    \eta_{28} 
    =\frac{q_1^2 \gam -\sqrt{\eta_1 \eta_9}}{q_1^2 \gam +\sqrt{\eta_1 \eta_9}}&\nonumber \\
    &
    \eta_{29}
    = \frac{q_2^2\gam - \sqrt{\eta_2 \eta_8} }{q_2^2\gam + \sqrt{\eta_2 \eta_8} }&
    &
    \eta_{30}
    = -\frac{q_1^2 +q_2^2 -2 \sqrt{\eta_2 \eta_9} }{q_1^2 +q_2^2 +2 \sqrt{\eta_2 \eta_9}} &\nonumber \\
    & 
    \eta_{31}
    =-\frac{q_1^2 +q_2^2 -2 \sqrt{\eta_1 \eta_8} }{q_1^2 +q_2^2 +2 \sqrt{\eta_1 \eta_8}} &
    &
    \eta_{32}
    = -\frac{q_1^2 -q_2^2 -2 \dom_1 \sqrt{\eta_1}}{q_1^2-q_2^2 +2 \dom_1 \sqrt{\eta_1} } &\nonumber \\
       & \eta_{33}
     = \frac{2\dom_1 (\dom_2^2-q_1^2)+(q_1^2+q_2^2)\dom_2 \gam - 2\sqrt{\eta_9 \eta_{16}}}{2\dom_1 (\dom_2^2-q_1^2)+(q_1^2+q_2^2)\dom_2 \gam + 2\sqrt{\eta_9 \eta_{16}}} 
    &
     & \eta_{34}
      = \frac{2\dom_2 (\dom_1^2-q_2^2)+(q_1^2+q_2^2)\dom_1 \gam - 2\sqrt{\eta_8 \eta_{16}}}{2\dom_2 (\dom_1^2-q_2^2)+(q_1^2+q_2^2)\dom_1 \gam+2\sqrt{\eta_8 \eta_{16}}} &
     \nonumber \\ 
     & \eta_{35}
     = \frac{
     q_1^2\dom_1 + q_2^2 (\dom_1 -2\dom_2 \gam)- 2\sqrt{\eta_2 \eta_{16}}
     }{
      q_1^2\dom_1 + q_2^2 (\dom_1 -2\dom_2 \gam)+ 2\sqrt{\eta_2 \eta_{16}}
     } &
     & \eta_{36}
     = \frac{
     q_2^2\dom_2 + q_1^2 (\dom_2 -2\dom_1 \gam)- 2\sqrt{\eta_1 \eta_{16}}
     }{
      q_2^2\dom_2 + q_1^2 (\dom_2 -2\dom_1 \gam)+ 2\sqrt{\eta_1 \eta_{16}}
     } & 
\end{align} 
\begin{align}
 &
    \eta_{37}
    = \frac{-q_1^2 -q_2^2 +2 \dom_1 \dom_2 \gam + (q_1^2 + q_2^2)\gam^2 - 2\sqrt{\eta_5 \eta_6 \eta_{16}}}{-q_1^2 -q_2^2 +2 \dom_1 \dom_2 \gam + (q_1^2 + q_2^2)\gam^2 + 2\sqrt{\eta_5 \eta_6 \eta_{16}}} \nonumber \\
     &
    \eta_{38}
    = \frac{2 q_2^2 \dom_2^2 - q_1^4 (\gam^2 -1)+q_1^2 (-2 \dom_1\dom_2 \gam + q_2^2 (\gam^2-1))-2q_1^2 \sqrt{\eta_5 \eta_6 \eta_{16}}}{2 q_2^2 \dom_2^2 - q_1^4 (\gam^2 -1)+q_1^2 (-2 \dom_1\dom_2 \gam + q_2^2 (\gam^2-1))+2q_1^2 \sqrt{\eta_5 \eta_6 \eta_{16}}} \nonumber \\
    & \eta_{39}
    = \frac{-2 q_2^2 \dom_2 (\dom_2-\dom_1 \gam)-2 q_1^2 \dom_1 (\dom_1-\dom_2 \gam)+(q_1^2-q_2^2)^2(\gam^2-1)-2 (q_1^2-q_2^2)\sqrt{\eta_5 \eta_6 \eta_{16}}}{-2 q_2^2 \dom_2 (\dom_2-\dom_1 \gam)-2 q_1^2 \dom_1 (\dom_1-\dom_2 \gam)+(q_1^2-q_2^2)^2(\gam^2-1)+2 (q_1^2-q_2^2)\sqrt{\eta_5 \eta_6 \eta_{16}}} \nonumber \\
    & \eta_{40}
    = 
    \frac{q_1^2 q_2^2 (\dom_2 + \dom_1 \gam) -q_1^2 \dom_1 (2 \dom_1 \dom_2 + q_1^2\gam)-q_2^4 \dom_2 -2 q_1^2 \dom_1 \sqrt{\eta_{16}}}{q_1^2 q_2^2 (\dom_2 + \dom_1 \gam) -q_1^2 \dom_1 (2 \dom_1 \dom_2 + q_1^2\gam)-q_2^4 \dom_2 +2 q_1^2 \dom_1 \sqrt{\eta_{16}}} \nonumber \\
    & \eta_{41}
    =
     \frac{q_1^2 q_2^2 (\dom_2 + \dom_1 \gam) -q_1^2 \dom_1 (2 \dom_1 \dom_2 + q_1^2\gam)-q_2^4 \dom_2 -2 q_2^2 \dom_2 \sqrt{\eta_{16}}}{q_1^2 q_2^2 (\dom_2 + \dom_1 \gam) -q_1^2 \dom_1 (2 \dom_1 \dom_2 + q_1^2\gam)-q_2^4 \dom_2 +2 q_2^2 \dom_2 \sqrt{\eta_{16}}} \nonumber \\
    & \eta_{42}
     = 
     \frac{q_1^2 \dom_1 (\dom_1 -\dom_2 \gam)+q_2^2 \dom_2 (\dom_2 -\dom_1 \gam)-2 \dom_1^2\dom_2^2+2\dom_1 \dom_2 \sqrt{\eta_{16}}}{q_1^2 \dom_1 (\dom_1 -\dom_2 \gam)+q_2^2 \dom_2 (\dom_2 -\dom_1 \gam)-2 \dom_1^2\dom_2^2-2\dom_1 \dom_2 \sqrt{\eta_{16}}}
\end{align}
The system of DEs in $\dd \log$ form is provided in the ancillary file \texttt{dlogDEs.m}. \\ 
In canonical form we expect only uniform transcendentality functions to appear at each order in $\epsilon$. Following reference~\cite{Bohnenblust:2023qmy}, boundary conditions can be fixed by solving the system of differential equations numerically via \texttt{AMFlow}~\cite{Liu:2022chg}. One can then analytically reconstruct the result order by order in $\epsilon$ via the \texttt{PSLQ}~algorithm~\cite{Duhr:2019tlz}, using an ansatz containing at order $n$ all the UT functions of weight $n$.
The result for the canonical MIs at $\cO\left(\epsilon^2\right)$ is:
\begin{align}
    \mathcal{G}_1
    & = 
    -\frac{\epsilon}{8 \pi }+\frac{\epsilon^2}{8 \pi } \left[ \log \left(\frac{\eta_3^2}{\pi}\right)-i \pi \right]+\cO\left(\epsilon^3\right)\ ,
    \\
    \mathcal{G}_2
    & = 
  -\frac{i \epsilon^2}{16}+\cO\left(\epsilon^3\right)\ ,
    \\
    \mathcal{G}_3
    & = 
    -\frac{\epsilon^2}{16 \pi } \left[\log \left(\frac{\eta_2}{\eta_{18}^2}\right)+i \pi \right]+\cO\left(\epsilon^3\right)\ ,
    \\
    \mathcal{G}_4
    & = 
   \frac{\epsilon}{16 \pi }+\frac{\epsilon^2}{16 \pi} \left[\log \left(\frac{4 \pi  \eta_{5} \eta_{6}}{\eta_4^2}\right)-i \pi \right]+\cO\left(\epsilon^3\right)\ ,
   \\
   \mathcal{G}_5
   & = 
   \frac{\epsilon}{16 \pi }+\frac{\epsilon^2}{16 \pi
   } \left[\log \left(\frac{4\pi \eta_5 \eta_6}{\eta_{19}^2\eta_3^2}\right)-i \pi \right]+\cO\left(\epsilon^3\right)\ ,
   \\
    \mathcal{G}_6
   & =
   \frac{\epsilon}{32 \pi }+\frac{\epsilon^2}{32
   \pi } \left[ \log \left(\frac{\eta_2^2\pi}{\eta_{1}^2\eta_3^2}\right)+i \pi \right]+\cO\left(\epsilon^3\right)\ ,
   \\
   \mathcal{G}_7
   & = 
   \frac{\epsilon}{32 \pi }+\frac{\epsilon^2}{32 \pi } \left[\log \left(  \frac{\pi}{\eta_4^2\eta_{19}^2}\right)+i \pi \right]+\cO\left(\epsilon^3\right)\ ,
   \\
   \mathcal{G}_8
   & = 
   \frac{\epsilon}{16 \pi }+\frac{\epsilon^2}{16 \pi
   } \left[ \log \left(\frac{4 \pi\eta_{4}^2}{\eta_1^2\eta_5 \eta_6}\right)+i \pi\right]+\cO\left(\epsilon^3\right)\ ,
   \\
   \mathcal{G}_9
   & = 
   \frac{\epsilon}{16 \pi }+\frac{\epsilon^2}{16 \pi } \left[\log \left(\frac{4 \pi \eta_3^2}{\eta_2^2 \eta_5\eta_6\eta_{19}^2}\right)+i
   \pi \right]+\cO\left(\epsilon^3\right)\ ,
   \\
   \mathcal{G}_{10}
   & = 
   \cO\left(\epsilon^3\right)\ ,
\end{align}
which agree with the integrals reported in \cite{Brandhuber:2023hhy,Herderschee:2023fxh,Caron-Huot:2023vxl,Bohnenblust:2023qmy}.

\bibliographystyle{JHEP.bst}
\bibliography{binary.bib}

\end{document}